\pdfoutput=1
\documentclass[12pt,letterpaper]{iopart}

\usepackage{graphicx}
\usepackage{subfigure}
\usepackage{caption}
\usepackage{siunitx}
\usepackage{placeins}
\usepackage{color}
\usepackage{standalone}
\usepackage{dcolumn}
\usepackage{bm}
\usepackage{microtype}
\usepackage{etoolbox}
\usepackage{amssymb}
\usepackage{accents}
\usepackage[normalem]{ulem}
\usepackage[dvipsnames]{xcolor}
\usepackage[colorlinks,urlcolor=NavyBlue,citecolor=NavyBlue,linkcolor=NavyBlue,pdfusetitle]{hyperref}
\usepackage[all]{hypcap}
\usepackage{enumerate}
\usepackage{enumitem}
\usepackage{gensymb}
\usepackage{comment}
\usepackage{tabularx,multirow}
\usepackage{array, makecell}
\usepackage{adjustbox}
\usepackage{cite}
\graphicspath{{./figs/}}

\newtoggle{commentsoff}
\newtoggle{highlight}

\ifdefined\nocomments
	\toggletrue{commentsoff}
\fi

\ifdefined\highlight
    \toggletrue{highlight}
\fi

\iftoggle{commentsoff}{
	\newcommand*{\tmp}[1]{}
	\newcommand*{\red}[1]{}
	\newcommand*{\ls}[1]{}
	\newcommand*{\owner}[1]{}
}{
	\newcommand*{\tmp}[1]{{\color{red} [{\bf Placeholder}: #1]}}
	\newcommand*{\red}[1]{{\color{red} #1}}
	\newcommand*{\ls}[1]{{\color{red} [{\bf LS}: #1]}}
	\newcommand*{\owner}[1]{{\color{blue} [{\bf Owner}: #1]}}
}

\iftoggle{highlight}{
	
}{
    
}

\newcommand{\dcc}{LIGO--P2100207}

\newcommand{\PreserveBackslash}[1]{\let\temp=\\#1\let\\=\temp}
\newcolumntype{C}[1]{>{\PreserveBackslash\centering}p{#1}}
\newcolumntype{R}[1]{>{\PreserveBackslash\raggedleft}p{#1}}
\newcolumntype{L}[1]{>{\PreserveBackslash\raggedright}p{#1}}

\begin{document}
\title[]{Characterization of systematic error in Advanced LIGO calibration in the second half of O3}

\author{
Ling~Sun$^{1,2}$, 
Evan~Goetz$^{3}$,
Jeffrey~S.~Kissel$^{4}$,
Joseph~Betzwieser$^{5}$,
Sudarshan~Karki$^{6}$,
Dripta~Bhattacharjee$^{6}$,
Pep~B.~Covas$^{10}$,
Laurence~E.~H.~Datrier$^{11}$, 
Shivaraj~Kandhasamy$^{12}$,
Yannick~K.~Lecoeuche$^{4}$,  
Gregory~Mendell$^{4}$,
Timesh~Mistry$^{13}$,
Ethan~Payne$^{14}$,
Richard~L.~Savage$^{4}$,
Aaron~Viets$^{7}$,
Madeline~Wade$^{8}$,
Alan~J.~Weinstein$^{1}$,
Stuart~Aston$^{5}$,  
Craig~Cahillane$^{1}$,
Jennifer~C.~Driggers$^{4}$,
Sheila~E.~Dwyer$^{4}$,
and Alexander~Urban$^{16}$}

\address {$^{1}$LIGO, California Institute of Technology, Pasadena, CA 91125, USA}
\address{$^{2}$OzGrav-ANU, Centre for Gravitational Astrophysics, College of Science, The Australian National University, ACT 2601, Australia}
\address {$^{3}$University of British Columbia, Vancouver, BC V6T 1Z4, Canada}
\address {$^{4}$LIGO Hanford Observatory, Richland, WA 99352, USA }
\address {$^{5}$LIGO Livingston Observatory, Livingston, LA 70754, USA}
\address {$^{6}$Institute of Multi-messenger Astrophysics and Cosmology, Missouri Institute of Science and Technology, Rolla, MO 65409, USA}
\address {$^{7}$Concordia University Wisconsin, 12800 N Lake Shore Dr, Mequon, WI 53097, USA }
\address {$^{8}$Kenyon College, Gambier, OH 43022, USA }
\address {$^{10}$Universitat de les Illes Balears, IAC3---IEEC, E-07122 Palma de Mallorca, Spain}
\address {$^{11}$SUPA, University of Glasgow, Glasgow G12 8QQ, UK}
\address {$^{12}$Inter-University Centre for Astronomy and Astrophysics, Pune 411007, India }
\address {$^{13}$The University of Sheffield, Sheffield S10 2TN, UK }
\address {$^{14}$OzGrav, School of Physics \& Astronomy, Monash University, Clayton 3800, Victoria, Australia }
\address {$^{15}$LIGO, Massachusetts Institute of Technology, Cambridge, MA 02139, USA }
\address {$^{16}$Louisiana State University, Baton Rouge, LA 70803, USA }

\ead{joseph.betzwieser@ligo.org}

\vspace{10pt}
\begin{indented}
\item[]\today~~
\end{indented}

\begin{abstract}
	We present the probability distribution of the systematic errors in the most accurate, high-latency version of the reconstructed dimensionless strain $h$, at the Hanford and Livingston LIGO detectors, used for gravitational-wave astrophysical analysis, including parameter estimation, in the last five months of the third observing run (O3B).
This work extends the results presented in Sun et. al (2020) \cite{Sun2020} for the first six months of the third observing run (O3A).
The complex-valued, frequency-dependent, and slowly time-varying systematic error (excursion from unity magnitude and zero phase) in O3B generally remains at a consistent level as in O3A, yet changes of detector configurations in O3B have introduced a non-negligible change in the frequency dependence of the error, leading to larger excursions from unity at some frequencies and/or during some observational periods; in some other periods the excursions are smaller than those in O3A.
For O3B, the upper limit on the systematic error and associated uncertainty is $11.29\%$ in magnitude and $9.18$ deg in phase (68\% confidence interval) in the most sensitive frequency band 20--2000 Hz. The systematic error alone is estimated at levels of $<2\%$ in magnitude and $\lesssim 4$ deg in phase.
These errors and uncertainties are dominated by the imperfect modeling of the frequency dependence of the detector response functions rather than the uncertainty in the absolute reference, the photon calibrators.

\end{abstract}

\section{Introduction}
\label{sec:introduction}
The Advanced Laser Interferometer Gravitational-Wave Observatory (Advanced LIGO) detectors~\cite{LIGO2014} and the Virgo Detector~\cite{Virgo2014} have directly observed transient gravitational waves from multiple compact binary coalescences in the first and second observing runs~\cite{GWTC-1, GWTC-2}.
After a series of instrument upgrades to further improve the sensitivity, e.g., replacing test masses and optics, increasing laser power, and adding squeezed light~\cite{buikema2020sensitivity}, the two LIGO detectors started the third observing run (O3), together with Virgo, on April 1st, 2019, and ended the first half of O3 (O3A) on Oct 1st, 2019.  
After completing a month long period of maintenance and upgrades, the second half of O3 (O3B) started on Nov 1st, 2019 and ran through Mar 27th, 2020. It ended about a month earlier than planned due to the COVID-19 stay-at-home orders.

In this paper, we describe changes to the estimation of the calibration systematic errors associated with improvement to the methods and responses to changes in the detectors during O3B. 
It is a supplement to Ref.~\cite{Sun2020} (which only discusses sources of systematic errors during O3A), and must be read in that context. 
This paper does not attempt to redefine all of the quantities defined in \cite{Sun2020}, and contains more technical details and jargon than would be appropriate for a stand-alone published article. 
Ref.~\cite{Sun2020} provides a full description of the calibration model and methods. Namely, structures of the parameterized control system model and methods for characterizing and combining the systematic error therein are essentially identical to those presented in \cite{Sun2020}. 
However, the errors found in the model during O3B are different than those in O3A, with some characterized using improved methods, and are thus described in this supplement.
Below, we briefly review the material from \cite{Sun2020} which is most relevant for understanding this supplement.

The Advanced LIGO detectors output a dimensionless strain $h$ (not the gravitational-wave strain), defined by the differential changes in arm length (DARM length) $\Delta L_{\rm free}$ divided by the average length of the arms $L$, 
\begin{equation}
	\label{eqn:strain}
	h = \frac{\Delta L_{\rm free} }{L}  = \frac{\Delta L_{\rm x}  -\Delta L_{\rm y} }{L}, 
\end{equation}
where $\Delta L_{\rm x}$ and $\Delta L_{\rm y}$ are the displacements in the two orthogonal arms, X and Y, respectively.
In the DARM feedback control loop, the residual DARM displacement $\Delta L_{\rm res}$ is converted by the sensing function $C$, to produce the digital output $d_{\rm err}$. 
The error signal is filtered through a set of digital filters $D$, creating the digital control signal, $d_{\rm ctrl}$ (i.e., $d_{\rm ctrl} = D d_{\rm err}$). 
The actuation function $A$, consisting of $A_i$ $(i = U, P, T)$ for the bottom three suspension stages, converts $d_{\rm ctrl}$ to the control displacement $-\Delta L_{\rm ctrl}$.
With this model of the control system, $\Delta L_{\rm free}$ is reconstructed as
\begin{equation}
	\label{eqn:displacement}
	\Delta L_{\rm free} = \Delta L_{\rm res} + \Delta L_{\rm ctrl} = \frac{1}{C^{\rm (model)}}  d_{\rm err} + A^{\rm (model)}  d_{\rm ctrl}.
\end{equation}
We can define a modeled response function, ${R}^{\rm (model)}$, given by
\begin{eqnarray}
	\label{eqn:response}
	{R}^{\rm (model)} = \frac{1+{A}^{\rm (model)} {D} {C}^{\rm (model)}}{{C}^{\rm (model)}},
\end{eqnarray}
such that 
\begin{equation}
	\label{eqn:dLfree}
	h = \frac {R^{\rm (model)}d_{\rm err}}{L}.
\end{equation}
See equations (1)--(4) and the full definition of functions $C$ and $A$ in section 2 of \cite{Sun2020}.

The frequency-dependent and time-dependent systematic error of the modeled response function ${R}^{\rm (model)}$, equivalent to the systematic error in estimated $h$, is defined by
\begin{equation}
	{\eta}_R \equiv \frac{{R}}{{R}^{\rm (model)}} = \frac{\delta {R}}{{R}^{\rm (model)}} + 1\,,
\end{equation}
where $\delta {R}/{R}^{\rm (model)} = {\eta}_R-1$ is the relative error in the response function as defined in~\cite{Cahillane2017}, and $R$ is the true response function.
The systematic error of the response function is represented by the deviation of $\eta_R$ from unity magnitude and zero phase.
The ultimate goal of accurate calibration is to keep the relative error as low as possible, i.e., ideally $\delta {R}=0$ across the entire frequency band at all times.

It is also useful to revisit the definitions of systematic error in individual components of $R^{\rm (model)}$ and how they contribute to $\eta_{R}$.
A systematic error in ${C}$, defined by ${\eta}_C \equiv {C}/{C}^{\rm (model)}$, will impact the response function systematic error as
\begin{equation}
	{\eta}_{R;C} = \frac{1}{{R}^{\rm (model)}} \left[\frac{1}{ {\eta}_C {C}^{\rm (model)}}  +{A}^{\rm (model)}{D} \right]\,.\label{eqn:etaC}
\end{equation}
Similarly, a systematic error in ${A}_i$ ($i=U,P,T$), defined by ${\eta}_{A_i} \equiv {A}_i/{A}_i^{\rm (model)}$, will impact the response function systemic error as
\begin{equation}
	{\eta}_{R;A_i} = \frac{1}{{R}^{\rm (model)}}  \left[  \frac{1}{{C}^{\rm (model)}} +  \left(  {\eta}_{A_i}  {A}_i^{\rm (model)} + \sum\limits_{j \neq i} {A}_j^{\rm (model)}\right) {D}\right]\,.\label{eqn:etaA}
\end{equation}
In Ref.~\cite{Sun2020}, the accuracy and precision of the detector response model, and equivalently the reconstructed $h$, are reported for O3A.
The complex-valued, frequency-dependent systematic error in the model (or equivalently in $h$) is presented in terms of the 68\% confidence interval contours of the error probability distribution estimated at any given time, denoted by $\eta_{R}(f;t)$~\cite{Sun2020}. 
We show the results of similar analyses here for O3B. 

The structure of this paper is as follows.  
We first quantify the systematic error probability distribution $\eta_{R}(f;t)$ in the detector response $R$ and equivalently in the estimate of strain $h$ as the ``final results'' for O3B in \sref{sec:results}, as in section 5 of \cite{Sun2020}. 
The provided error estimates pertain to the most accurate and final version of calibrated strain generated in O3B.
Sections~\ref{sec:pcal} through \ref{sec:gpr} discuss the enhancements, changes, and updated inputs of the models that lead to these final results.
Finally, in \sref{sec:conclusion}, we conclude and discuss the O3 results in terms of the systematic error impact on current and future gravitational wave analyses.

\section{Systematic error final results}
\label{sec:results}
In this section, we quantify the combined systematic error and uncertainty $\eta_{R}(f;t)$ over time, present in the modeled detector response $R^{\rm (model)}$ and thus in the data stream of dimensionless strain $h$ used for astrophysical analysis. 
We use the same numerical method as described in section 5.1 of \cite{Sun2020} to obtain the frequency-dependent probability distribution of $\eta_{R}$ at a given time.
As in section 5.2 of \cite{Sun2020}, the results are described by the statistical percentile contours, the so-called ``epoch distribution,'' of all the systematic error estimates at discrete times $t_{k}$ with hourly cadence, i.e., all $\eta_{R}(f;t_k)$, within a given epoch of the detector configuration.

\begin{table}[!tbh]
	\centering
	\caption{\label{tab:C01_results}O3A and O3B calibration epochs and the maximum $1\sigma$ and median excursions of response from unity magnitude and zero phase compared to the maximum a posteriori (MAP) response function ${R}_{\rm MAP}$ (see details in \cite{Sun2020}), in the most sensitive frequency band 20--2000~Hz. The maximum median values represent the best estimate of the systematic error bounds over 20--2000~Hz.}
	\setlength{\tabcolsep}{5pt}
	\begin{tabular}{lllll}
		\br
		Hanford epoch & Max $1\sigma$  & Max $1\sigma$  & Max median & Max median\\
		& magnitude [\%] & phase [deg] & magnitude [\%] & phase [deg]\\
		\mr
		(O3A a) Mar 28--Jun 11 & 6.96  & 3.79 & 1.58& 0.86 \\
		(O3A b) Jun 11--Aug 28 & 4.11  & 2.34  &1.15&0.92\\
		(O3A c) Aug 28--Oct 1 & 3.33  & 1.53  &1.42&1.00\\
		(O3B a) Nov 1--Jan 14 & 3.89 & 2.02 & 1.19 & 0.77\\
		(O3B b) Jan 14--Feb 11 & 3.42 & 1.83 & 1.20 & 0.82\\
		(O3B c) Feb 11--Mar 16 & 4.56 & 2.33 & 2.00 & 1.07\\
		(O3B d) Mar 16--Mar 27 & 11.29 & 9.18 & 1.97 & 4.32\\
		\br
		Livingston epoch & Max $1\sigma$  & Max $1\sigma$  & Max median & Max median\\
		& magnitude [\%] & phase [deg] & magnitude [\%] & phase [deg]\\
		\mr
		(O3A a) Mar 28--Jun 11 & 6.37  & 3.49  &1.13&1.59\\
		(O3A b) Jun 11--Oct 1 & 5.99  & 3.68  &1.09&2.09\\
		(O3B a) Nov 1--Jan 14 & 9.58 & 6.60 & 0.80 & 0.52\\
		(O3B b) Jan 14--Mar 27 & 8.63 & 5.98 & 0.61 & 0.52\\ 
		\br
	\end{tabular}
\end{table}

As described in section 5.2 of \cite{Sun2020}, the systematic error and uncertainty estimate for a given epoch across a given frequency band is quoted by two numbers (one for magnitude and the other for phase) for brevity, which indicate the maximum excursions from unity magnitude and zero phase (i.e., no systematic error) in that band. 
These numbers are listed in \tref{tab:C01_results} for the most sensitive frequency band of 20--2000~Hz.
We include the O3A epochs for completeness and comparison.
\Tref{tab:O3Bepoch} lists the epochs in O3B for both Hanford and Livingston detectors and the detector configuration changes that define the boundaries associated with each epoch. Note that at Hanford, epoch O3B b (starting on January 11, 2020) is created so that the data delivery is aligned with the Livingston configuration change. Although there is no detector configuration change for Hanford from O3B~a to O3B~b, the resulting numbers in \tref{tab:C01_results} are different for these two epochs. This is because more measurements are included in O3B~b such that the uncertainties are better constrained.

\begin{table}[!tbh]
	\centering
	\caption{\label{tab:O3Bepoch}O3B calibration epochs and the main changes in each epoch. The configuration changes leading to the epoch boundaries are discussed in more detail in sections \ref{sec:lho} and \ref{sec:llo}.}
	\setlength{\tabcolsep}{6pt}
	\begin{tabular}{ll}
		\br
		Hanford epoch & Changes  \\
		\mr
		(O3B a) Nov 1--Jan 14 & Start of O3B   \\
		(O3B b) Jan 14--Feb 11 & Start of the second data interval, chosen to align\\
		& with the LLO epoch\\
		(O3B c) Feb 11--Mar 16 &  Electrostatic driver current limit resistors replaced \\
		& with Safe High Voltage (SHV) barrels at the end \\
		& test mass on the X arm \\
		(O3B d) Mar 16--Mar 27 & OMC whitening chassis configuration changed \\
		\br
		Livingston epoch & Changes \\
		\mr
		(O3B a) Nov 1--Jan 14 & Start of O3B  \\
		(O3B b) Jan 14--Mar 27 & Adjusted the gain in the TST actuator due to a \\
		& 30\% drift \\
		\br
	\end{tabular}
\end{table}

In the left panels of figures~~\ref{fig:h1_results1}--\ref{fig:l1_results}, 
the white curves indicate the estimated frequency-dependent systematic error for each epoch. 
The 68\%, 95\%, and 99\% confidence intervals of the $1\sigma$ uncertainty boundaries in the epoch distributions are shown as dark, moderate, and light shaded regions, respectively. 
These epoch distributions quantify the time-dependent variation of the combined uncertainty and systematic error bounds over the entire epoch. 
The absolute values of the white curve (median) and 68\% boundaries in the left panels are plotted on the right. 
(See detailed description for figures~16 and 17 in \cite{Sun2020}.)
The features of these plots are discussed below.

\begin{figure}[!tbh]  \ContinuedFloat*	
        \centering
        \subfigure[]
        {
        	\label{fig:h1_O3B_chunk1}
        	\scalebox{0.225}{\includegraphics{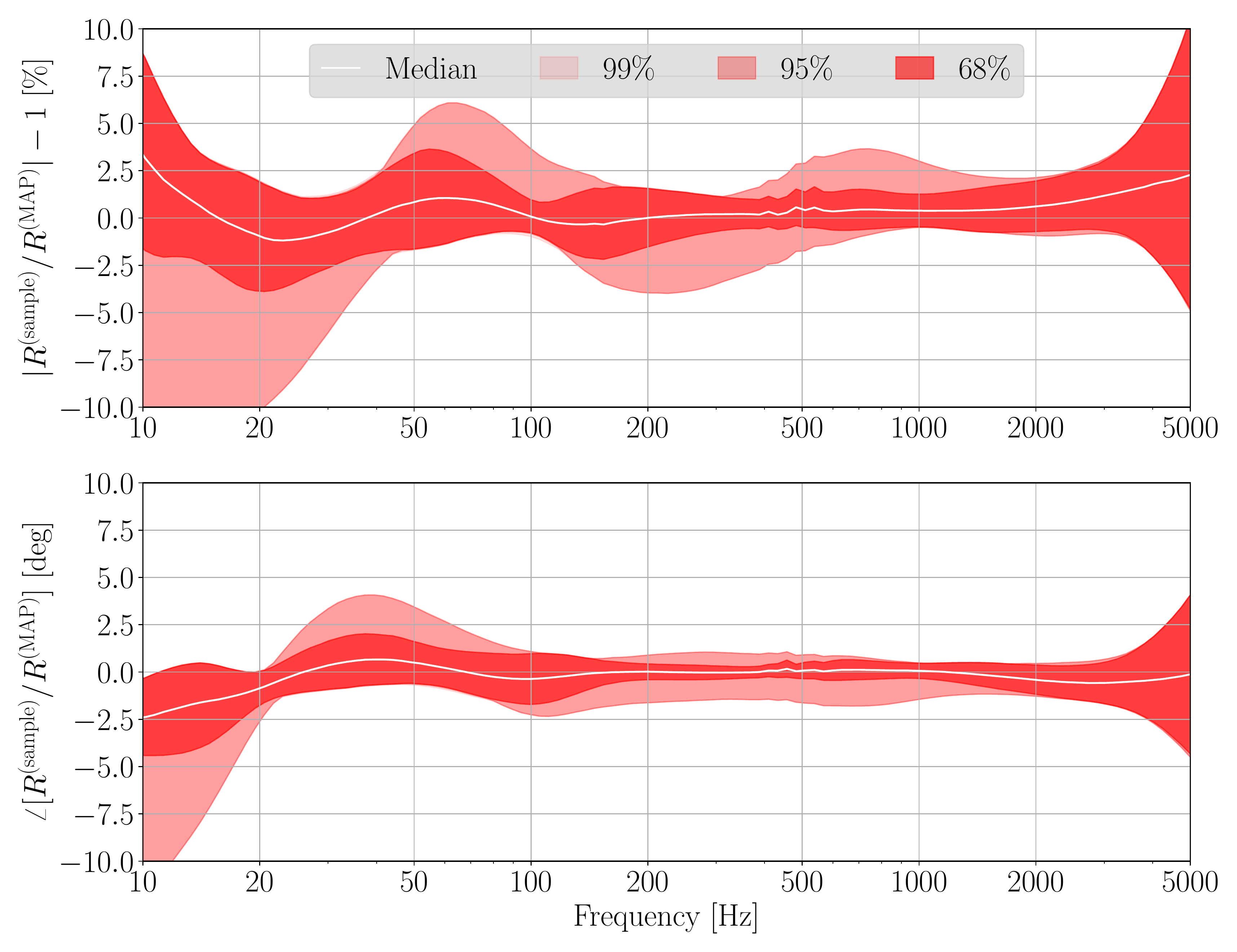}}
        	\scalebox{0.225}{\includegraphics{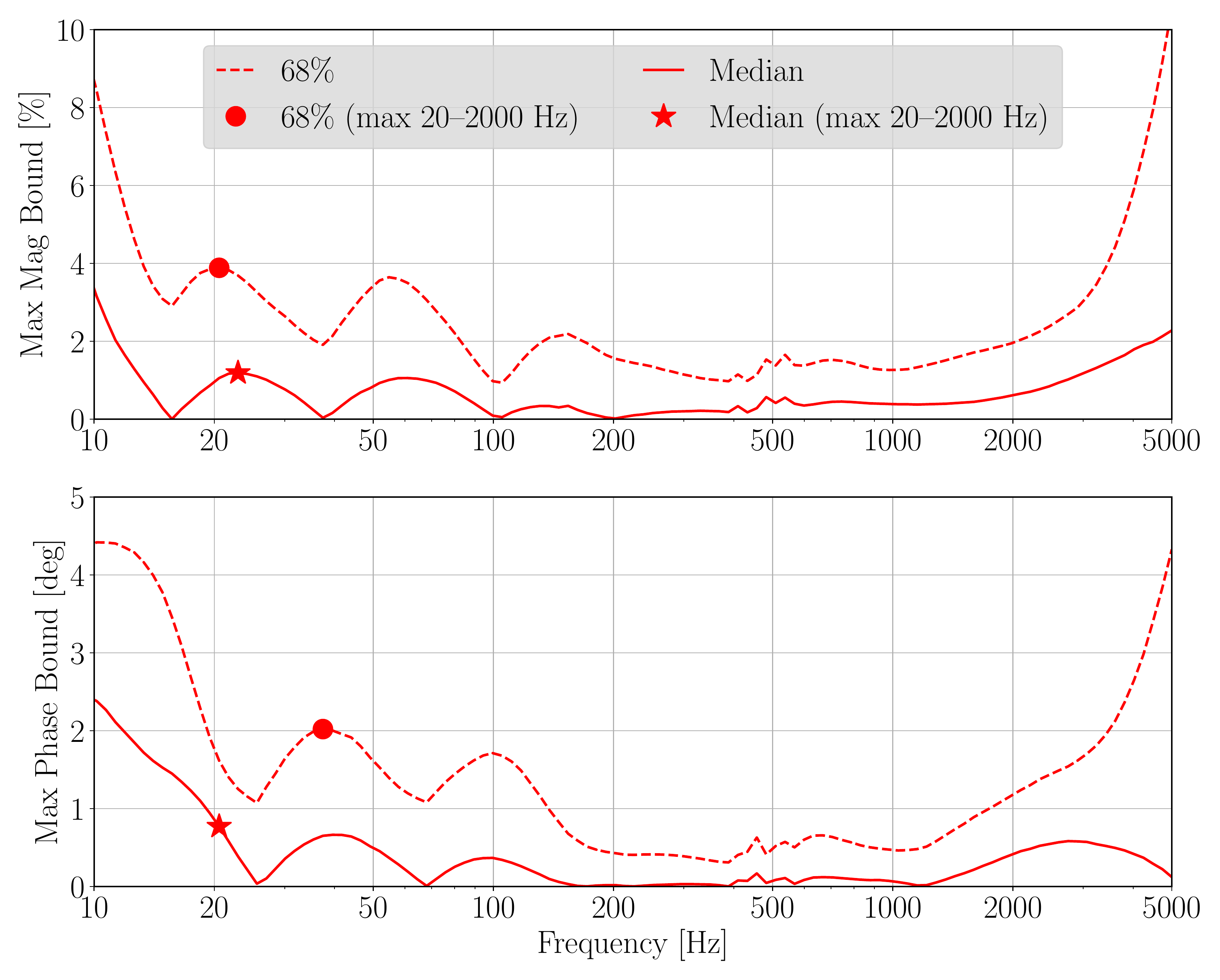}}
        }
        \subfigure[]
        {
                \label{fig:h1_O3B_chunk2a}
                \scalebox{0.225}{\includegraphics{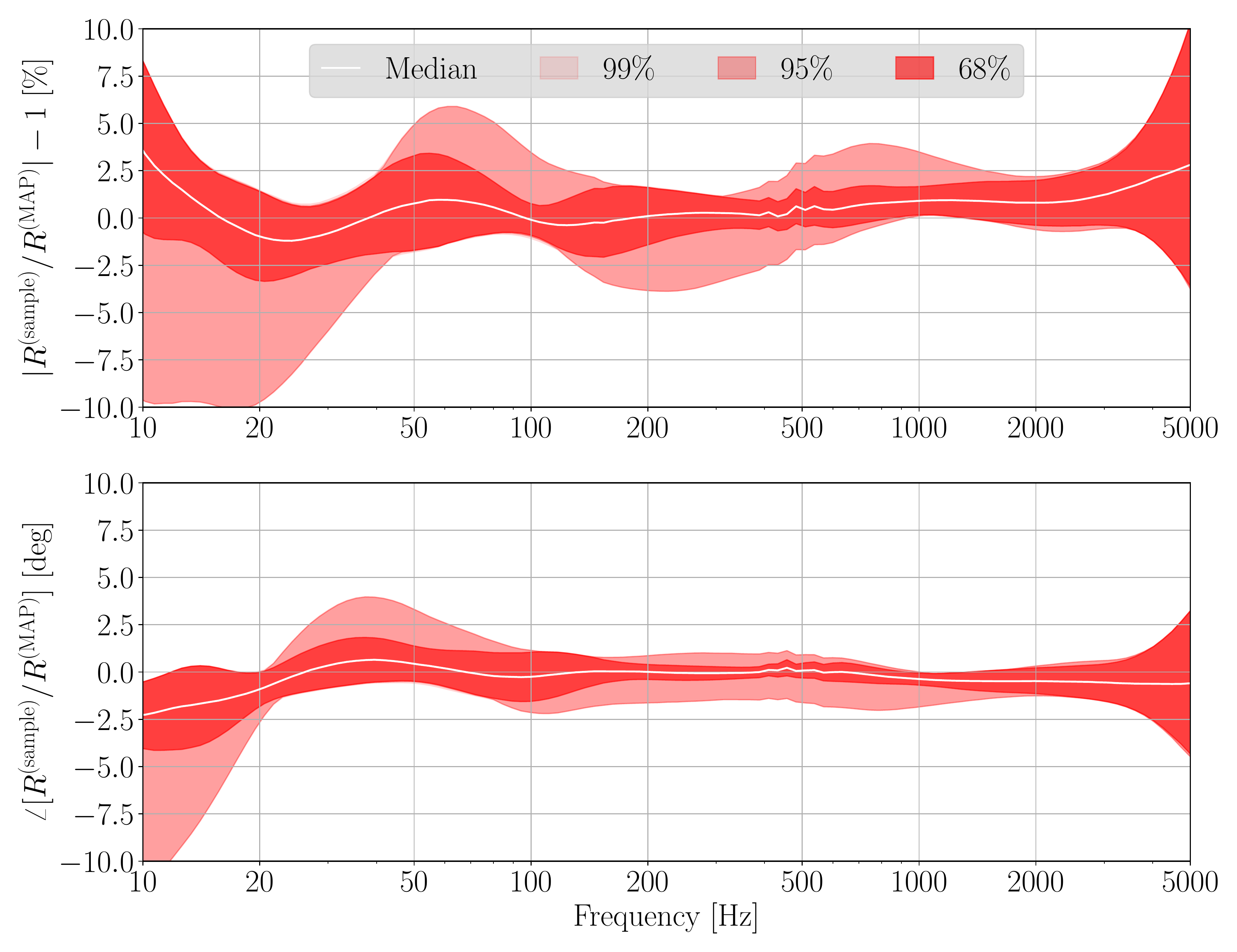}}
                \scalebox{0.225}{\includegraphics{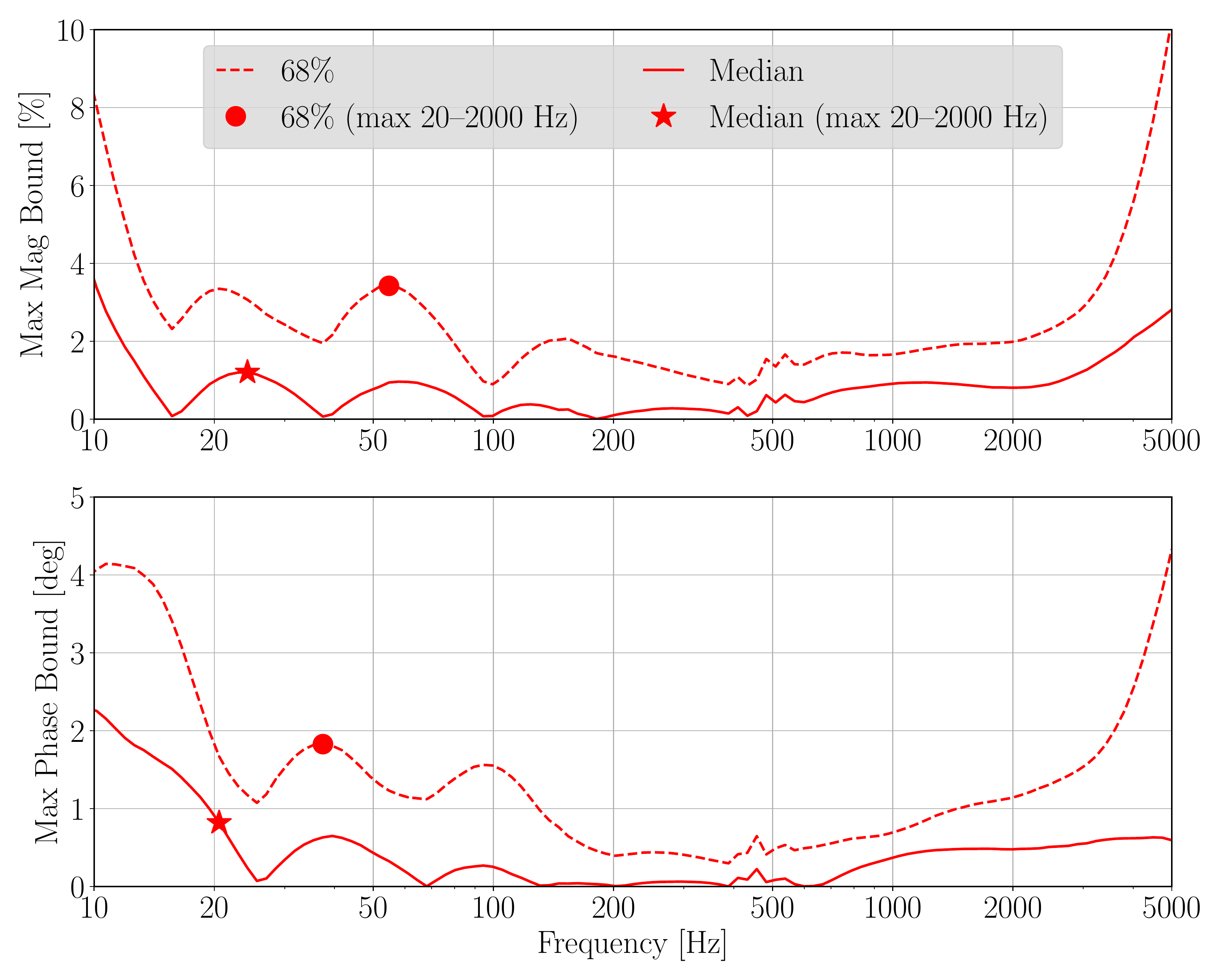}}
        }
        \caption[]{Variation of the combined systematic error and uncertainty (left) and the maximum bounds (right) for Hanford. The two subfigures correspond to Hanford O3B epochs (a)--(b) in \tref{tab:C01_results}. 
        	The top and bottom panels of each subfigure show the frequency dependent excursions of response from unity magnitude and zero phase compared to ${R}_{\rm MAP}$, respectively. The percentiles are obtained from all the hourly evaluated ${\eta}_R(f;t_k)$ over each epoch. 
        	In the left panels, the colors represent $1\sigma$ uncertainty for 68\%, 95\%, and 99\% of the run time, as indicated in the legend. The 95\% and 99\% percentiles overlap each other but deviate from the 68\% percentile (see details in text). The white curve indicates the median excursion. 
        	The absolute values of the boundaries (median and 68\%) in the left panels are plotted on the right. The star and dot markers indicate the median and $1\sigma$ maximum excursions in the frequency band 20--2000~Hz, respectively.}
        \label{fig:h1_results1}
\end{figure}

\begin{figure}[!tbh]
	\centering
	\subfigure[]
	{
		\label{fig:h1_O3B_chunk2b}
		\scalebox{0.225}{\includegraphics{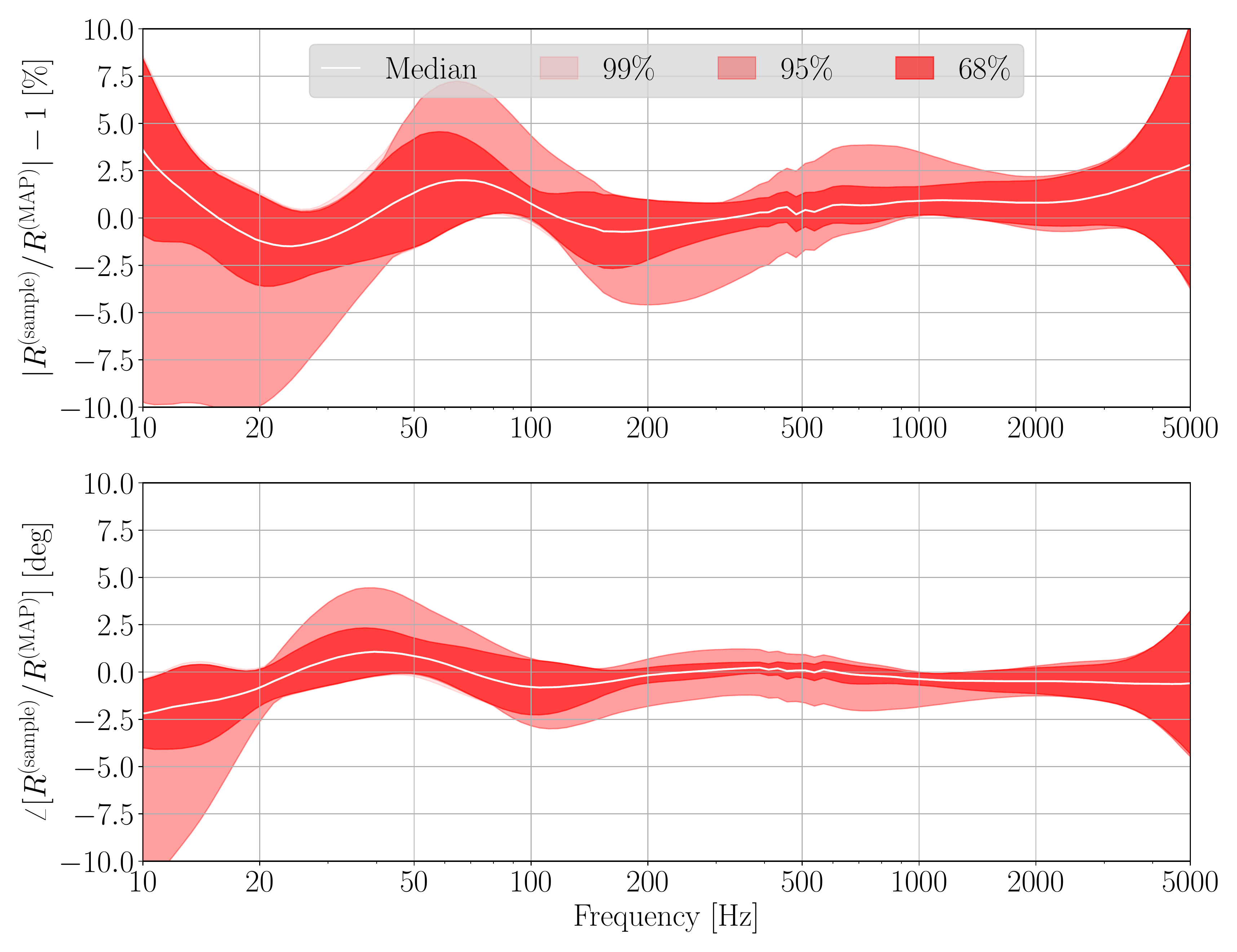}}
		\scalebox{0.225}{\includegraphics{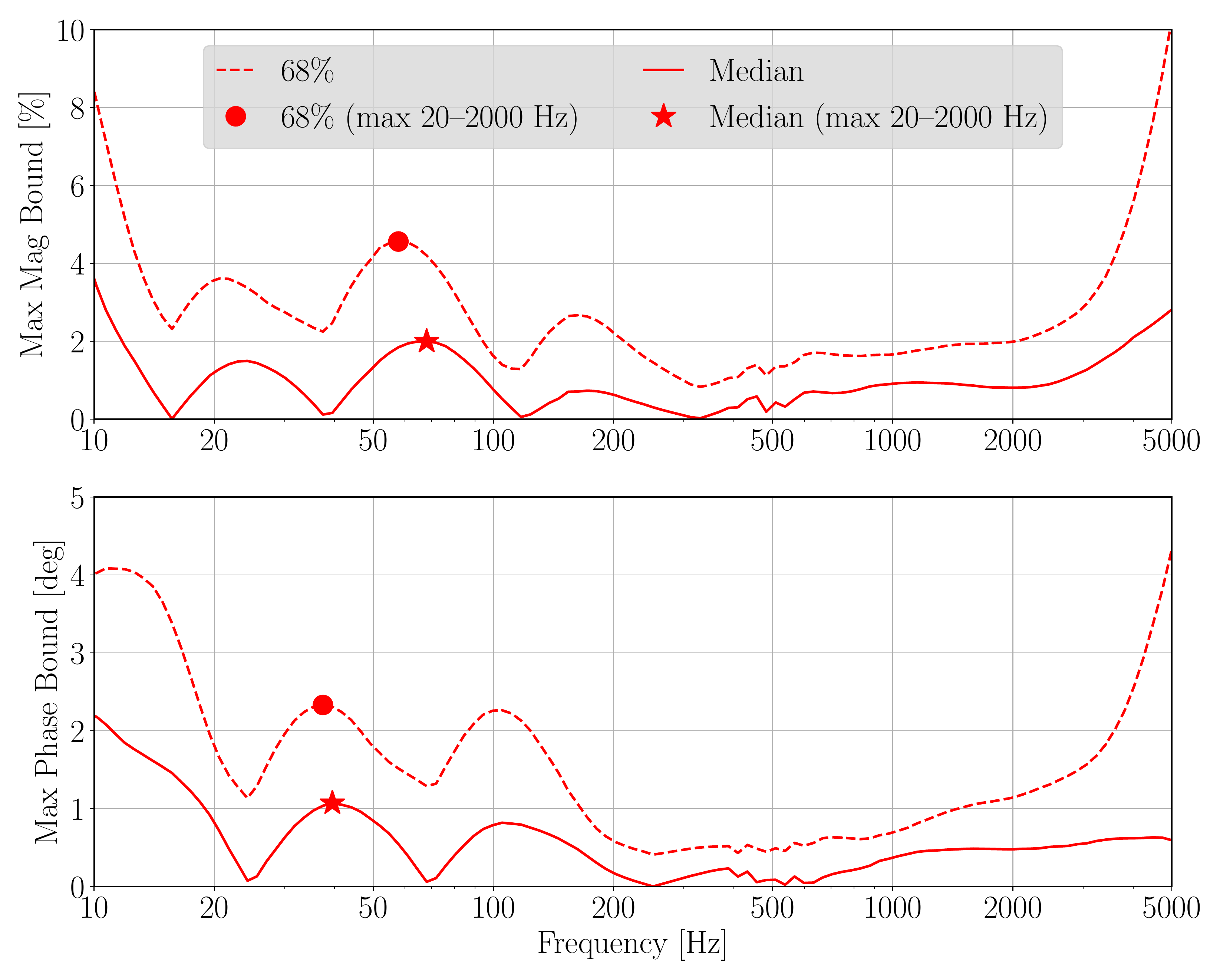}}
	}
	\subfigure[]
	{
		\label{fig:h1_O3B_chunk2c}
		\scalebox{0.225}{\includegraphics{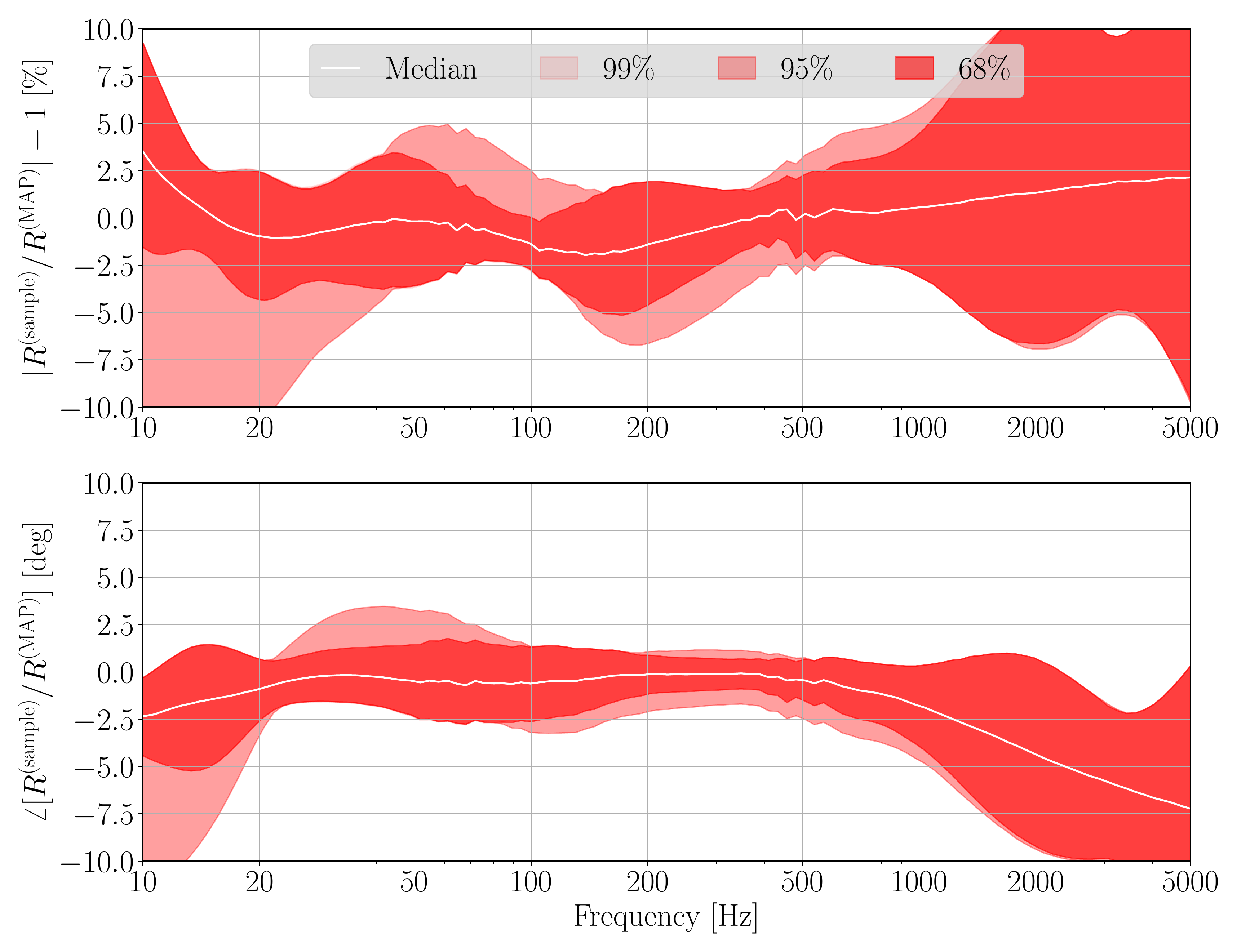}}
		\scalebox{0.225}{\includegraphics{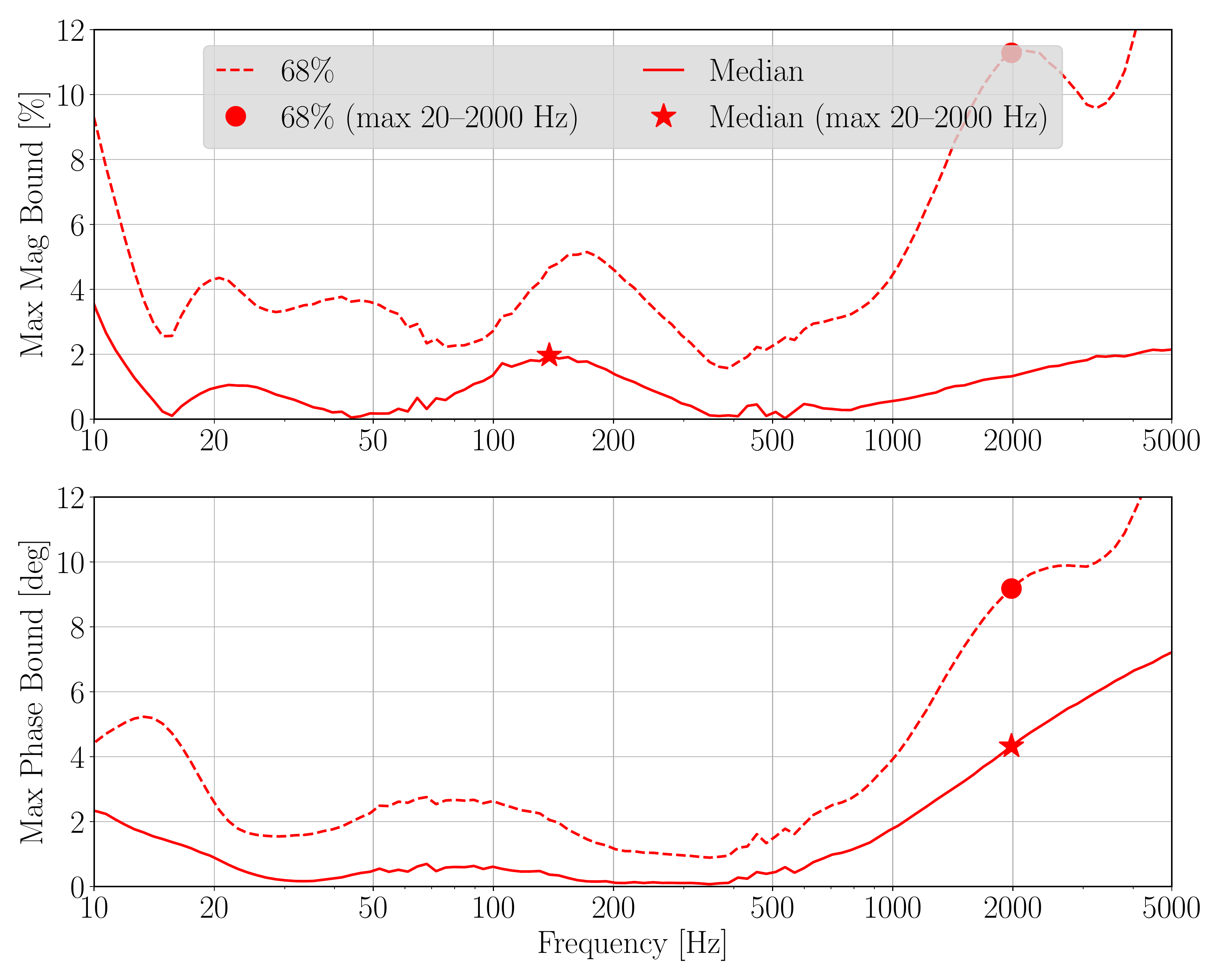}}
	}
	\caption[]{(Continued.)--Variation of the combined systematic error and uncertainty (left) and the maximum bounds (right) for Hanford. The two subfigures correspond to Hanford O3B epochs (c)--(d) in \tref{tab:C01_results}. }
	\label{fig:h1_results2}
\end{figure}

\begin{figure}[!tbh]
        \centering
        \subfigure[]
        {
                \label{fig:l1_O3B_chunk1}
                \scalebox{0.225}{\includegraphics{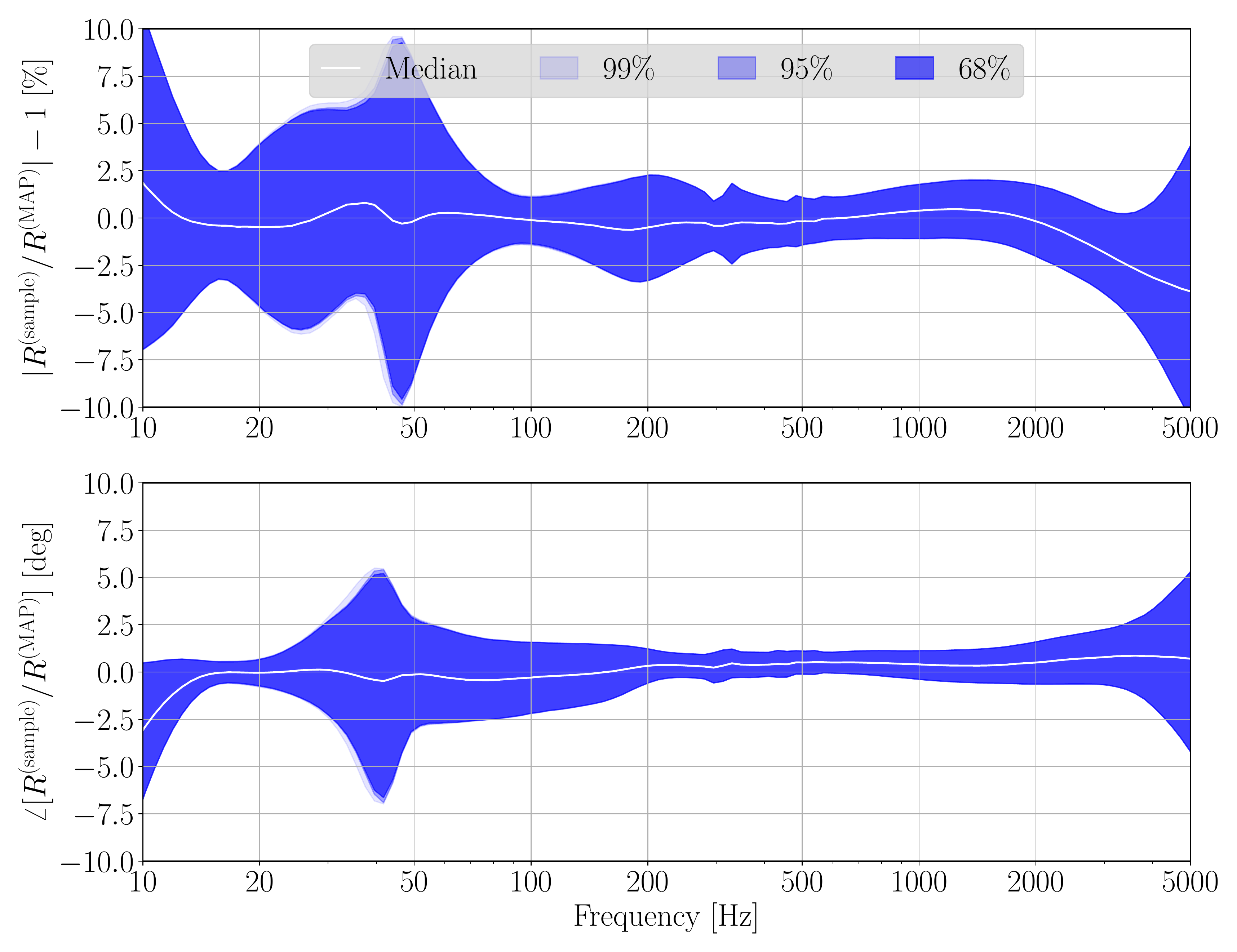}}
                \scalebox{0.225}{\includegraphics{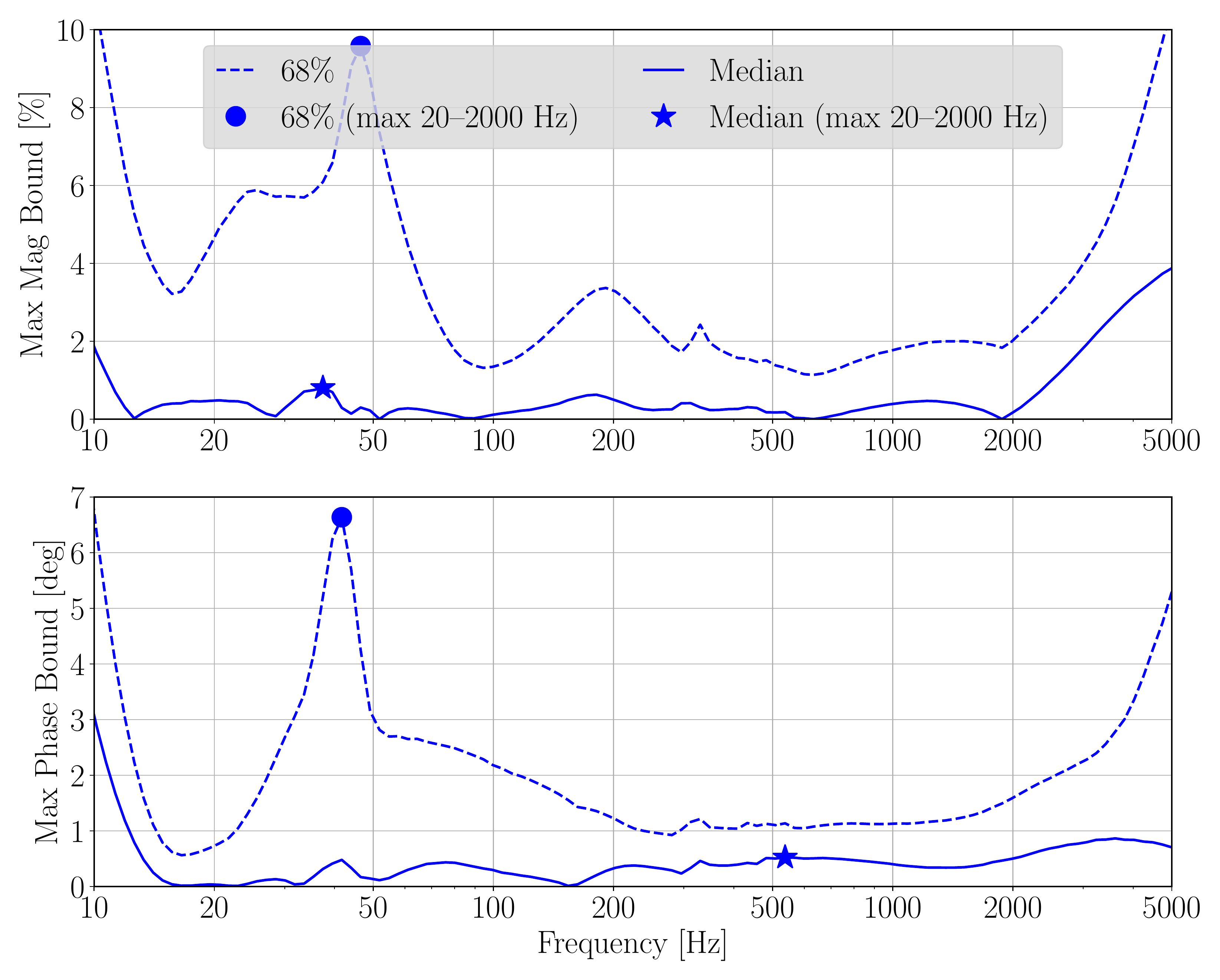}}
        }
        \subfigure[]
        {
                \label{fig:l1_O3B_chunk2}
                \scalebox{0.225}{\includegraphics{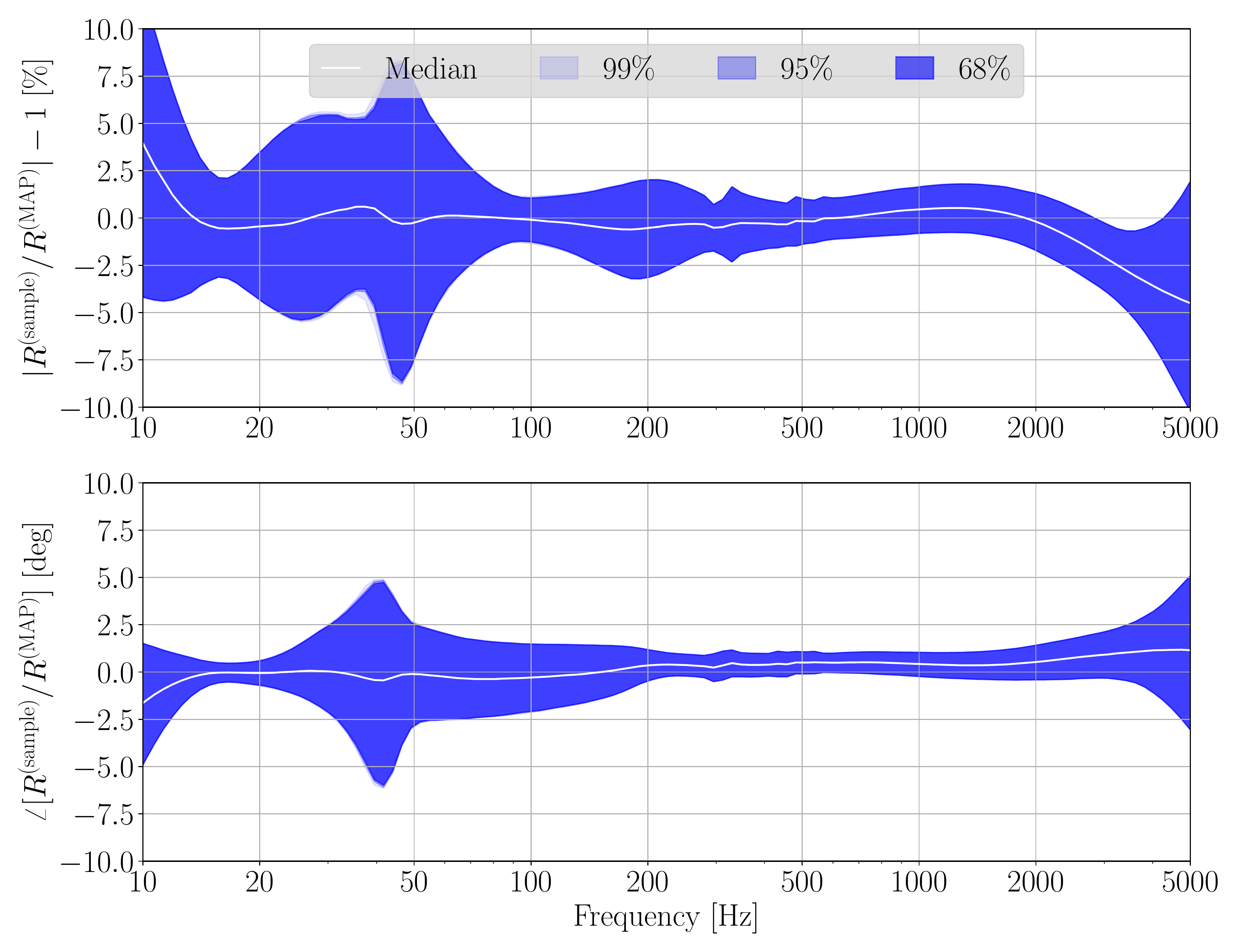}}
                \scalebox{0.225}{\includegraphics{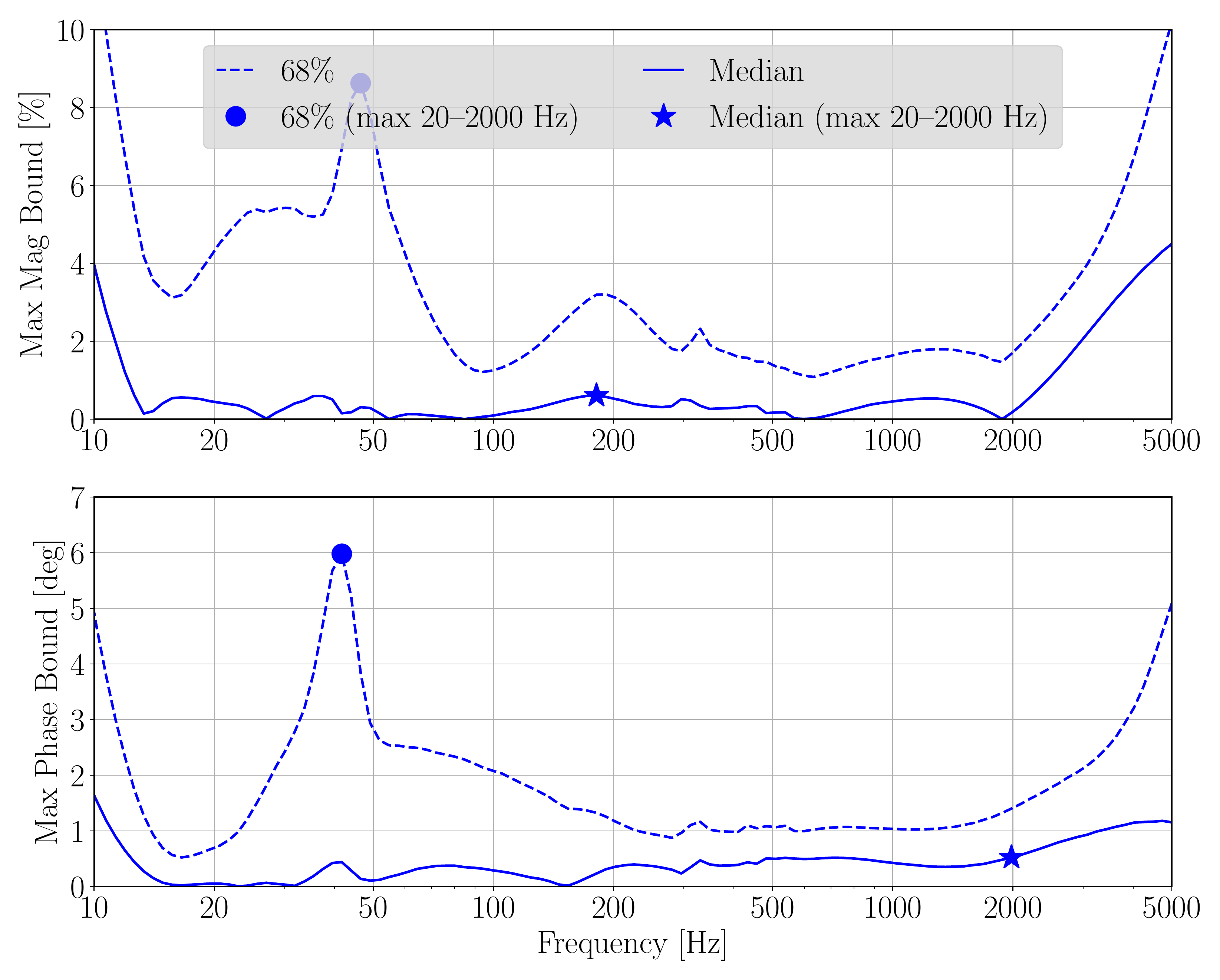}}
        }
        \caption[]{Variation of the combined systematic error and uncertainty (left) and the maximum bounds (right) for Livingston. The two subfigures correspond to Livingston O3B epochs (a)--(b) in \tref{tab:C01_results}.
                The top and bottom panels of each subfigure show the frequency dependent excursions of response from unity magnitude and zero phase compared to ${R}_{\rm MAP}$, respectively. The percentiles are obtained from all the hourly evaluated ${\eta}_R(f;t_k)$ over each epoch.
                In the left panels, the colors represent $1\sigma$ uncertainty for 68\%, 95\%, and 99\% of the run time, as indicated in the legend. These percentiles generally overlap with each other, indicating that the variation of the uncertainty bounds is negligible at Livingston. The white curve indicates the median excursion. 
                The absolute values of the boundaries (median and 68\%) in the left panels are plotted on the right. The star and dot markers indicate the median and $1\sigma$ maximum excursions in the frequency band 20--2000~Hz, respectively.}
        \label{fig:l1_results}
\end{figure}

At Hanford, the variation in the uncertainty bounds in O3B is larger than O3A (c.f. figure~16 in \cite{Sun2020}), evidenced by that the 95\% and 99\% percentiles overlap each other but deviate significantly from the 68\% interval.
This is a result of a better representation of the deficiencies in the detector sensing function model.
This new representation is discussed in more detail in \sref{sec:thermal} and summarized here as follows. 
As the laser power input into the detector increases on the way for the detector to achieve its nominal low-noise state, and the optics are far from thermal equilibrium, the sensing function below $\sim$30~Hz is further distorted beyond the other problems discussed in section 4.2 of \cite{Sun2020}.
Defined by the thermal time constant of the optics, this systematic error is the worst immediately after the power increase, and exponentially diminishes to negligible within two hours when the detector eventually reaches thermal equilibrium.
The detector noise performance is unaffected by this error; gravitational wave detections are possible as soon as the nominal low-noise state is achieved, prior to achieving thermal equilibrium.

This effect is dependent on the mirror's absorption. It only showed up at Hanford as the Hanford detector happened to have a worse mirror. The Livingston detector did not suffer from such effect.
While this effect is present at the Hanford detector throughout O3, the impact is different between O3A and O3B due to the configuration changes in between. We have adopted different approaches to estimate and quantify the error in O3A and O3B. 
In O3A, this error is qualitatively covered by a conservative uncertainty estimate at low frequencies in the sensing function (see details in section 4.3 of \cite{Sun2020}).
In O3B, we develop a method to quantitatively estimate this error and include it in the overall systematic error and uncertainty estimate of the detector response.  
A detailed description is given in \sref{sec:thermal}.

As in O3A (see section 5.2 in \cite{Sun2020}), the systematic error is estimated at discrete times $t_{k}$ with 1-hr cadence (from the start of each epoch) to inform the error time-dependent variation over each epoch. At each discrete time $t_k$, we first check the uncertainty of the injected sinusoidal excitations, i.e., ``calibration lines'' (see section 4.1 in \cite{Sun2020}).
If the uncertainty (calculated using Eq. (19) in \cite{Sun2020}) is sufficiently low, indicating that the detector is in a state with sufficiently low noise to detect gravitational waves, an estimate of the error is computed for that given time. Otherwise, the error estimate is skipped for that $t_{k}$ since the data at that time are not qualified for astrophysical analyses.
In O3B, we apply an additional, fixed, frequency-dependent error calculated as described in \sref{sec:thermal} to each estimate $\eta_{R}(f;t_k)$ with $t_k$ falling within an hour after the laser power increases and the automated lock acquisition system~\cite{rollins2016distributed} reports that it has achieved the nominal low-noise state.
Using this method, roughly 10\% of the Hanford detector hourly estimates have this extra error applied throughout O3B (consistent with the detector duty cycle~\cite{buikema2020sensitivity}), affecting all of the Hanford epochs in \tref{tab:O3Bepoch}. Therefore the 95\% and 99\% percentile curves report the enlarged error for the $\sim 10\%$ of time in all Hanford epochs.

\Fref{fig:h1_O3B_chunk2c} shows relatively larger uncertainties at Hanford for Epoch O3B d.
This is because only a limited number of measurements were taken during the final 11-day epoch, after the detector configuration changed on March 16, 2020 (see details in \sref{sec:electronics}).
As such, the estimate of the unknown systematic error and uncertainty via Gaussian Process Regression (GPR) in the sensing function is poorly constrained, leading to increased uncertainty of the overall response function systematic error estimate.

At Livingston, the elevated contribution of the test mass (TST) actuator to the response function at $\sim50$~Hz (see detailed discussion in \sref{sec:lloESDgain}) continues to result in the relatively larger uncertainty around 50~Hz in O3B (see \fref{fig:l1_results}).
This is essentially unchanged from O3A and also indicates slightly higher overall combined systematic error and uncertainty (see figure 17 in \cite{Sun2020}).
The variation of the overall uncertainty bounds remains negligible at Livingston in O3B (i.e., the 68\%, 95\%, and 99\% percentiles generally overlap with each other in both O3A and O3B).

\section{Update to Pcal absolute reference}
\label{sec:pcal}
Photon calibrators (Pcal), which independently use photon radiation pressure to produce strain within the detector~\cite{Karki2016,Pcalpaper-P2000113}, are the primary absolute reference used to validate the estimates of $h$ itself as well as the error and uncertainty of the detector response.  Calibration of the displacement fiducials provided by the Pcal systems is realized by control system variables in code running on the computers that convert the signals from the Pcal receiver-side sensors at the interferometer end stations to end test mass displacement. The  amplitude (in meters) of a displacement (x($\omega$)) induced by periodic Pcal forces at angular frequency $\omega$ is given by X/$\omega^2$  multiplied by receiver-side power sensor signal, in digital counts (for details of how the displacement factors are implemented, see \cite{Pcalpaper-P2000113, Pcal-G1501518}).  
Variables derived from the displacement factor X are used in the front-end code to calibrate the Pcal end station power sensor outputs. Their values were updated at the beginning of the O3 observing run, based on the limited data available at the time for the various components of the Pcal calibrations. They were updated at the end of the O3A portion of the observing run, then again at the end of the O3B portion of the run, taking into account both additional measurements of the relevant components of the Pcal calibrations and improvements in the methods for estimating uncertainties.  Improvements in the methods used to estimate Pcal uncertainties, as well as the details of the Pcal systematic error and uncertainties for the Hanford and Livingston detector during the O3 run can be found in \cite{Pcalpaper-P2000113,Pcal-T2100067, Pcal-T2100219}. 


The relevant dates and the associated Pcal displacement factors updated in the front end code in O3 for both end stations at both the Hanford and Livingston detectors are listed in \tref{tab:FCDC}.  The table summarizes the results of analyses captured in \cite{Pcal-T2100219} for the Hanford detector and in \cite{Pcal-T2100067} for the Livingston detector. 


If the raw Pcal signals were to be re-calibrated after the end of the O3 run, the values in the ``All of O3 run" rows of the tables would be used for the whole run.  However, given that the data have been calibrated in ``chunks" during the run, discrepancies between the ``All of O3 run'' displacement factor and the factors calibrating the Pcal signals during the intervals listed in the tables result in incorrectly calibrated data. 
The final column in the tables lists values for $\eta_{\rm Pcal}$, the multiplicative correction factors for the Pcal fiducials, i.e., the ratio between the correct Pcal fiducials and the estimated values used in each O3 interval due to errors (poorly or mistakenly formed estimates) in Pcal displacement factors.
Because these corrections factors are not applied when the data are calibrated, they indicate the Pcal-related relative errors remaining in the reconstructed strain data.
These $\eta_{\rm Pcal}$ factors together with their associated 1-$\sigma$ uncertainty $U_{\rm rel}$ are taken into account when numerically estimating $\eta_R$ (see section 5.1 in \cite{Sun2020}), and thus the Pcal systematic error uncertainties are included in the results presented in \sref{sec:results}.

\begin{table}[!tbh]
\centering
\caption{\label{tab:FCDC} Pcal calibration factors during the O3 run.  The first column lists the interferometer end station, the second column the dates for the relevant interval, the third column the displacement factors, and their estimated relative uncertainties (1-$\sigma$) in the fourth column; the last column lists the Pcal multiplicative correction factors by which the displacement factors in each interval should be multiplied to take advantage of all of the information and improvements in analysis methods available at the end of the observing run.} 
	\setlength{\tabcolsep}{4pt}
	\begin{tabular}{c c c c c}
    \br
    End & Front end   & X (displ. factor) & $U_{\rm rel}$ & $\eta_{\rm Pcal}$  \\
    Station & code interval  & Units: m/(s$^2$ct) & ($\%$) & \\
    \br
    Hanford  &$3/26/19:10/31/19$ & \num{1.5673e-14}  & 0.54 & 0.9960 \\
    X-end & $10/31/19:11/11/19$  & \num{1.5721e-14}  & 0.54 & 0.9929 \\
    & $11/11/19:4/1/20$ & \num{1.5711e-14}  & 0.54 & 0.9935 \\
    & All of O3 run & \num{1.5610e-14}  & 0.41 & NA \\
    \mr
    Hanford & $3/26/19:10/31/19$ & \num{1.5781e-14}  & 0.54 & 1.0018 \\
    Y-end & $10/31/19:11/11/19$ & \num{1.5840e-14}  & 0.54 & 0.9980 \\
    & $11/11/19:4/1/20$ & \num{1.5830e-14} & 0.54 & 0.9986 \\
    & All of O3 run & \num{1.5808e-14}  & 0.41 & NA \\ 
    \mr
    Livingston & $3/26/19:11/1/19$ & \num{1.6433e-14} & 0.54 & 1.0018 \\
     X-end &  $11/1/19:4/1/20$ & \num{1.6519e-14} & 0.54 & 0.9966 \\
    & All of O3 run & \num{1.6463e-14} & 0.44 & NA \\
    \mr
    Livingston  & $3/26/19:11/1/19$ & \num{1.6143e-14} & 0.54 & 1.0037 \\
    Y-end& $11/1/19:4/1/20$ & \num{1.6194e-14} & 0.54 & 1.0005 \\
    & All of O3 run & \num{1.6203e-14} & 0.44 & NA \\
    \br
	\end{tabular}
\end{table}

\section{Hanford detector changes}
\label{sec:lho}
\subsection{Thermalization error handling}
\label{sec:thermal}
Throughout O3, the low-frequency ($\lesssim 30$~Hz) response of the Hanford detector displayed time-dependence within the first $\sim$2~hours of `observation-ready' low-noise operation. 
We attribute this time-dependence to a complicated interaction between at least the following three physical issues. 
First, the response of the detector changes as the interferometer optics converge toward thermal equilibrium after the input laser power is brought up to the level needed for observation. The timescales and effects of this process are complicated by point-like defects in the optical coatings referred to as ``point absorbers''~\cite{Brooks2020}. 
Second, in order to avoid the most impactful point defects near the geometric center of the optics, the global angular control system was used to position the laser beam spots on the arm cavity optics away from the geometric center of the optics, which introduces unintended mixing of global cavity length and angle control loops via the lever arm created by the distance between the spot position and the geometric center (and the center of rotation) of the optic.
Third, the error signal for microscopic length control of the signal recycling cavity is distorted by the changes of the modulated optical power available within the cavity, which deviates from design itself as a result of the point defects.
This deviation of the error signal causes the control system to settle on a sub-optimal zero point for the error signal, leading to a position offset between the controlled cavity length and the optimal physical length of the cavity, and therefore detuning the signal recycling cavity with respect to the arm cavities.
It proves to be a challenge to derive a model of the collection of these effects from first principles, especially as they are suspected to be mixed together. 
Thus, we instead employ empirical models based on measurements of the response resulted from whatever combination of effects that takes place.

In O3A, the control system was adjusted (namely an artificial digital time-invariant offset was applied to the signal recycling cavity length), such that we saw the phenomenological ``spring-like'' response dominated sensing functions, with the square of the spring pole frequency $f_{s}^{2} > 0$ (as those in O1 and O2) or $f_{s}^{2} < 0$ (new in O3), or in a more complicated case, further altered by a dominant angular cross-coupling effect (see detailed discussion in section 4.2 of \cite{Sun2020}).
We derive an empirical model in O3A by performing a GPR on the collection of all standard ``swept-sine'' measurements of the sensing function, including those taken both within the $\sim2$-hr thermalization periods and after the detector is thermalized, but deliberately exclude measurements taken below 20~Hz, which vary significantly depending on the thermalization stages and other effects mentioned above.
Also, in the GPR process in O3A, we intentionally use agnostic priors and a length scale hyperparameter larger than that used in O3B (see \sref{sec:gpr}), such that the uncertainty estimate of the sensing response at low frequencies $\lesssim 20$~Hz is conservative enough to cover the unknown effects (see discussion in section 4.3 of \cite{Sun2020}). 

\begin{figure}[tbh!]
	\centering
	\subfigure[]
	{
		\label{fig:therm1}
		\scalebox{0.36}{\includegraphics{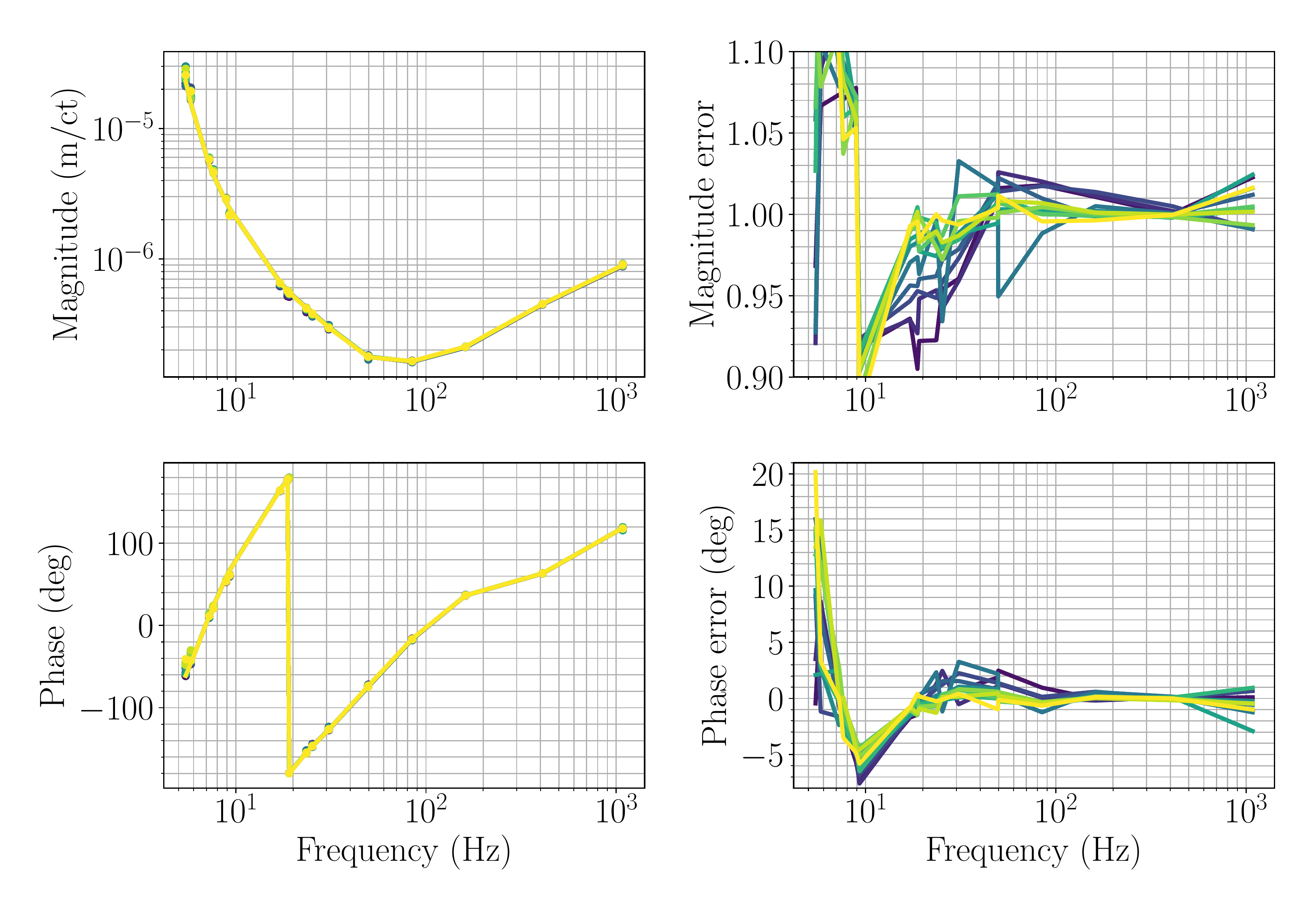}}
	}
	\subfigure[]
	{
		\label{fig:therm2}
		\scalebox{0.36}{\includegraphics{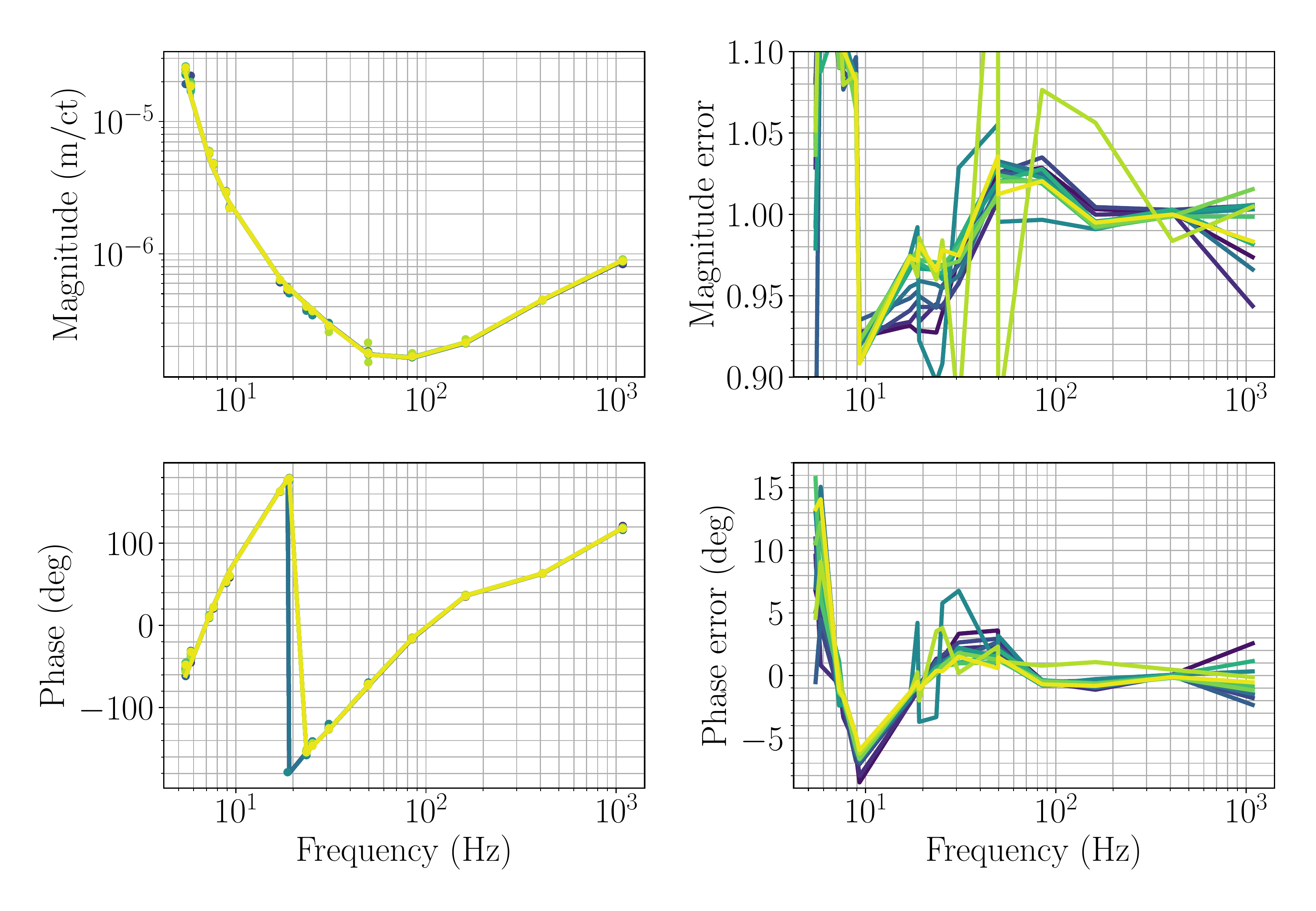}}
	}
	\caption[]{Measured response functions and errors from the two sets of measurements taken on (a) December 17, 2019 and (b) February 27, 2020 at Hanford. Measurements at the start are in darkest blue when the detector first enters low-noise configuration, while those in the lightest yellow are at the end as the detector has thermalized.}
	\label{fig:therm}
\end{figure}

In O3B, the fixed control offset applied to the signal recycling cavity length has been readjusted in such a way that the response is consistently dominated by ``spring-like'' response with $f_{s}^{2} > 0$ during the thermalization, and then evolves to having only a small amount of distortion throughout the rest of each observation-ready stretch (the minor distortion left is believed to come from the residual length-angle cross-coupling). 
In this way, the problem of distorted response from thermalization and that from unwanted control-loop cross-coupling are deemed separable and can thus be more accurately represented.
To quantify the remaining unknown frequency-dependent errors after thermalization, we use the same GPR method as in O3A (as described in \cite{Sun2020}), with a few updates to the GPR parameters that are further discussed in \sref{sec:gpr}. 
In this section, we describe the method for quantifying the errors during the first two hours of thermalization in O3B, and how it is integrated into the systematic error estimate of the overall response function.


Rather than using the standard ``swept-sine" measurements to characterize the sensing function as described in \cite{Sun2020}, we instead use the Pcal system to add several additional sinusoidal calibration lines, and monitor the response function, via equation (13) in \cite{Sun2020}, constantly for several hours, before, during, and after the detector enters low-noise operation (the data collected during this study are excluded from the strain product).
At each injected line frequency, the magnitude and phase of response function is computed at 2-min intervals to track the thermalization over time.
Analysis parameters [40~s per fast Fourier transform (FFT), Hann windowing, 50\% overlap; equating to 5 averages per measurement] are chosen to balance time resolution and signal-to-noise ratio of the measurement lines in each FFT.
This special measurement was repeated twice on two separate occasions, several months apart to verify the reproducibility of this effect.

These two sets of measurements are shown in \fref{fig:therm}. For each set, at each measurement time and frequency, the results are compared against the expected response based on the reference model, i.e., obtaining a direct measurement of the frequency-dependent $\eta_{R}$ as a function of time.
As the detector thermalizes, the early data points (blue curves) slowly evolve to the thermalized state (yellow curves).
Both measurements show a similar trend in response: right after the observation-ready low-noise state is achieved, the systematic error can be as large as 8\% in magnitude at 20~Hz and diminishes over a 2-hr period to only a small amount (the phase change is negligible throughout the 2-hr period).
Note that below 10~Hz and above 1~kHz, the actuation strength of the Pcal system is not strong enough to obtain sufficient signal-to-noise ratio and coherence of the measurement data points above detector noise on the short timescales tracking this effect.  
As such, the data points in these frequency ranges are not representative of the systematic error under investigation.
However, the measured detector response to the otherwise constantly injected calibration lines at 410~Hz and 1083~Hz (obtained with FFTs over much longer timescales) indicates that the distortion from this thermalization effect is entirely negligible above 1~kHz. 

\begin{figure}[!tbh]
	\centering
    \scalebox{0.7}{\includegraphics{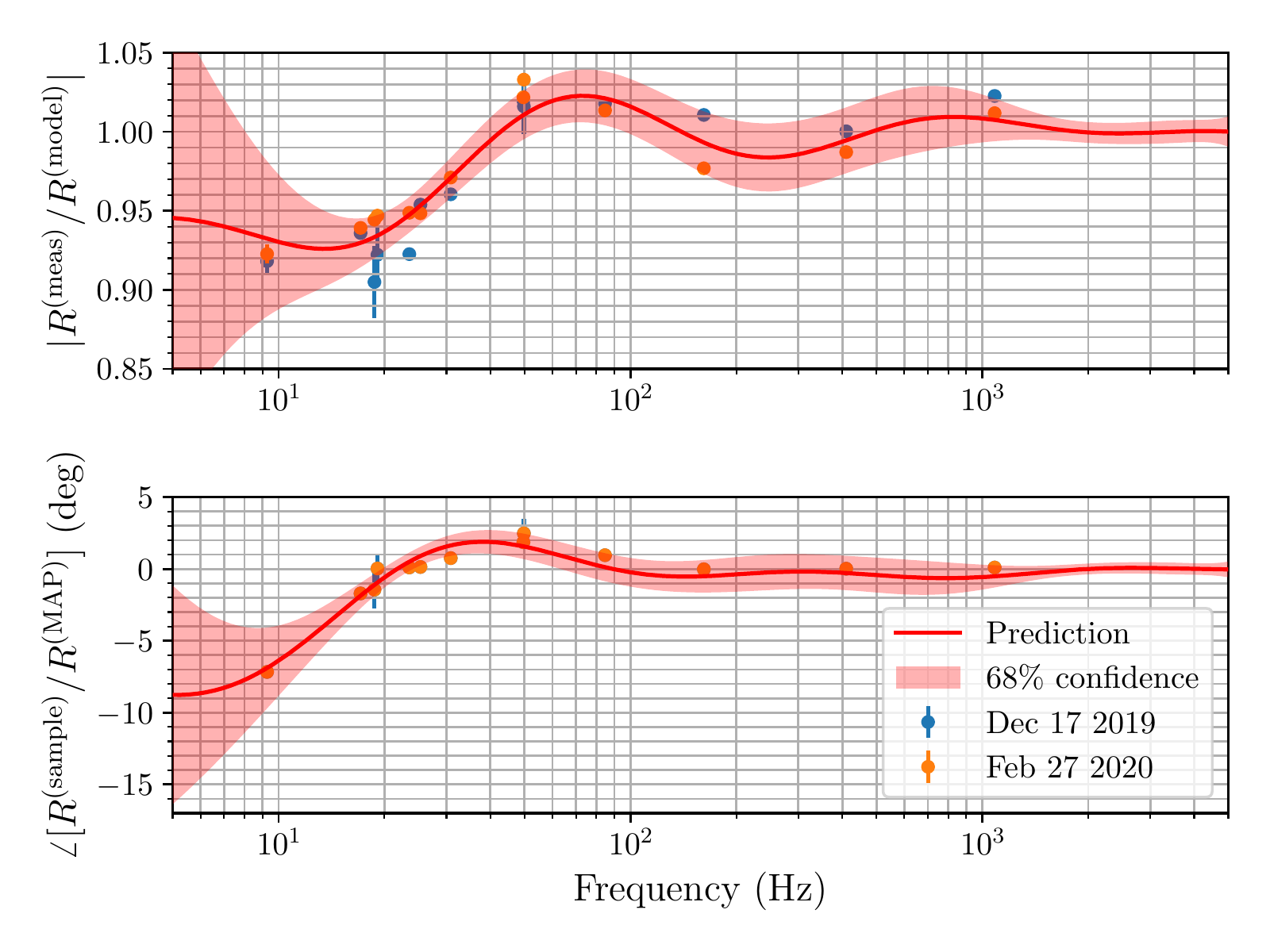}}
	\caption[]{Systematic error and uncertainty estimate of the Hanford response function right after the observation-ready low-noise state is achieved. Data points in two different colors indicate the measurements taken on two different days. The thick red curve is the maximum a postiori estimate of the frequency-dependent systematic error following a GPR applied to the two sets of data points. The shaded region indicates the 68\% confidence interval on the systematic error.}
	\label{fig:therm_gpr}
\end{figure}

After confirming that the error is reproducible (as expected), we create an empirical model of the directly-measured response function error, $\eta_{R}$, with a single GPR fit of the two measurements taken right after the observation-ready low-noise state is achieved (see \fref{fig:therm_gpr}).
Five artificial data points are added in the fit from 1.5~kHz to 5~kHz at unity magnitude and zero phase (where there is no impact from the low-frequency response changes), in order to obtain desired constraints in the GPR fitting above $\sim1$~kHz, where the signal-to-noise ratio of the measurement data points is not good enough or no data are available in this special study.
Additionally, data points below 9~Hz are excluded from the fit due to the low signal-to-noise ratio at low frequencies.

With this GPR fit, we have a static representation of the worst-case systematic error and associated uncertainty from this thermalization effect alone. We must then apply it to the the systematic error estimate of the overall response function only when this error is present, and derive the distribution of $\eta_{R}(f;t_{k})$, where $t_k$ can be a time with or without being impacted by this error.
The thermalization error is the worst right after the laser is powered up, and then smoothly, exponentially decays to a negligible level over $\sim 2$~hrs.
However, the overall error and uncertainty $\eta_{R}(f;t_{k})$ is estimated at discrete times $t_k$ with an 1-hr cadence (see details in \sref{sec:results} and section 5 of \cite{Sun2020}). 
As such, when calculating $\eta_{R}(f;t_{k})$ for an entire epoch ($t_{k}$ takes discrete time values with 1-hr cadence), where the detector configuration remains unchanged, we check within the most recent hour of data to see if the Hanford detector is both 1) in nominal low-noise and 2) at its nominal high-laser-power configuration.
If any of the two status values for low-noise and high-laser-power is not nominal within the most recent hour, it indicates that the detector is still thermalizing and impacted by the error, and thus the thermalization GPR fit (\fref{fig:therm_gpr}) is applied.
While this application does not perfectly capture the continuous evolution of the error, we argue that it sufficiently represents the discretized effect that``turns on'' for an hour (when it is still significant) and then ``turns off'' afterword (when it becomes negligible).
The results of this application are reflected in the statistics collecting all the hourly estimates over the whole observing time (see figures~\ref{fig:h1_results1} and \ref{fig:h1_results2}).

\subsection{Electronics changes}
\label{sec:electronics}
As the observing run progresses, non-Gaussian, non-stationary noise, and/or ``glitches'' are often identified in the data without a clear source or physical mechanism from which they arise. 
To aid noise mitigation efforts, the configuration of some individual component of a given detector control system is changed to judge whether that component is creating these excess noise sources. 
In O3B, at the Hanford detector, two of these configuration changes adversely affected the DARM control loop, and the problems were identified and the impact quantified only after the observing run ended.
After a complete study, the data are divided into distinct periods defined by the times of the configuration changes, and an updated level of systematic error in the response function is created for each period.
Sections \ref{sec:actuationelectronics-uim} and \ref{sec:actuationelectronics-tst} cover the effects of two changes to the quadruple pendulum actuator in O3B: 
(1) a change in configuration of the switchable low-pass filtering of the analog driver in the upper intermediate mass stage (UIM), and (2) a configuration change in the resistance along the signal path of the test mass stage (TST) actuation driver electronics outside of the vacuum enclosure. \Sref{sec:sensingelectronics} covers the effects of a change in the analog whitening electronics in the gravitational-wave readout photodiodes. For more details about suspension system signal processing electronics and a better explanation of the terms used in this section, see Refs.~\cite{carbone2012sensors,aston2012update}.

\subsubsection{UIM switchable analog filter change (between November 27 and December 3, 2019)\\}
\label{sec:actuationelectronics-uim}

The UIM coil drivers are a collection of analog circuits which drive current through the four voice-coil actuators on the UIM actuator stage of every quadruple pendulum (see figure (2) of~\cite{Sun2020}). 
The driver has switchable frequency response to allow for switching between higher-range mode during lock acquisition of the detector, and low-noise mode after the detector is under control. 
Each of the switchable frequency responses is assigned to a state number. In the case of the UIM driver, ``State 1'' is the highest range state; ``State 4'' is the lowest noise state; ``State 2'' and ``State 3'' are intermediate range/noise states in between.
To reduce the complexity of the up-stream, global control filter design, the response of any analog state of the driver is inverted, or ``compensated,'' in the digital system. 
The compensation filters are typically informed by either first-principle calculations of the expected response, or fits of direct measurements of the analog system response. 
Though the desire/intent is for the product of the analog response and the digital compensation response to be unity, there may remain some frequency-dependent error in the product of the analog circuit and its compensation filter. This remaining error is the frequency-dependent systematic error $\eta_{A_{U}}$ on the UIM actuation function $A_{U}$.\footnote{Note that early in O3B, we identified a systematic error in the response function model caused by an unaccounted for collection of sharp resonant features in the mechanical/dynamical force-to-displacement transfer function of the UIM stage. As further detailed in \cite{LHOalog55399}, this error in the UIM force-to-displacement transfer function was present until January 2020 in the low-latency data stream, and was fixed in the most accurate, high-latency calibrated data stream of O3B used for the most accurate astrophysical parameter estimation. But it remains as an uncompensated systematic error in the most accurate Hanford O3A data product; see section 4.4 (d) in \cite{Sun2020}.}

\begin{figure}[!tbh]
	\centering
	\includegraphics[width=0.8\textwidth]{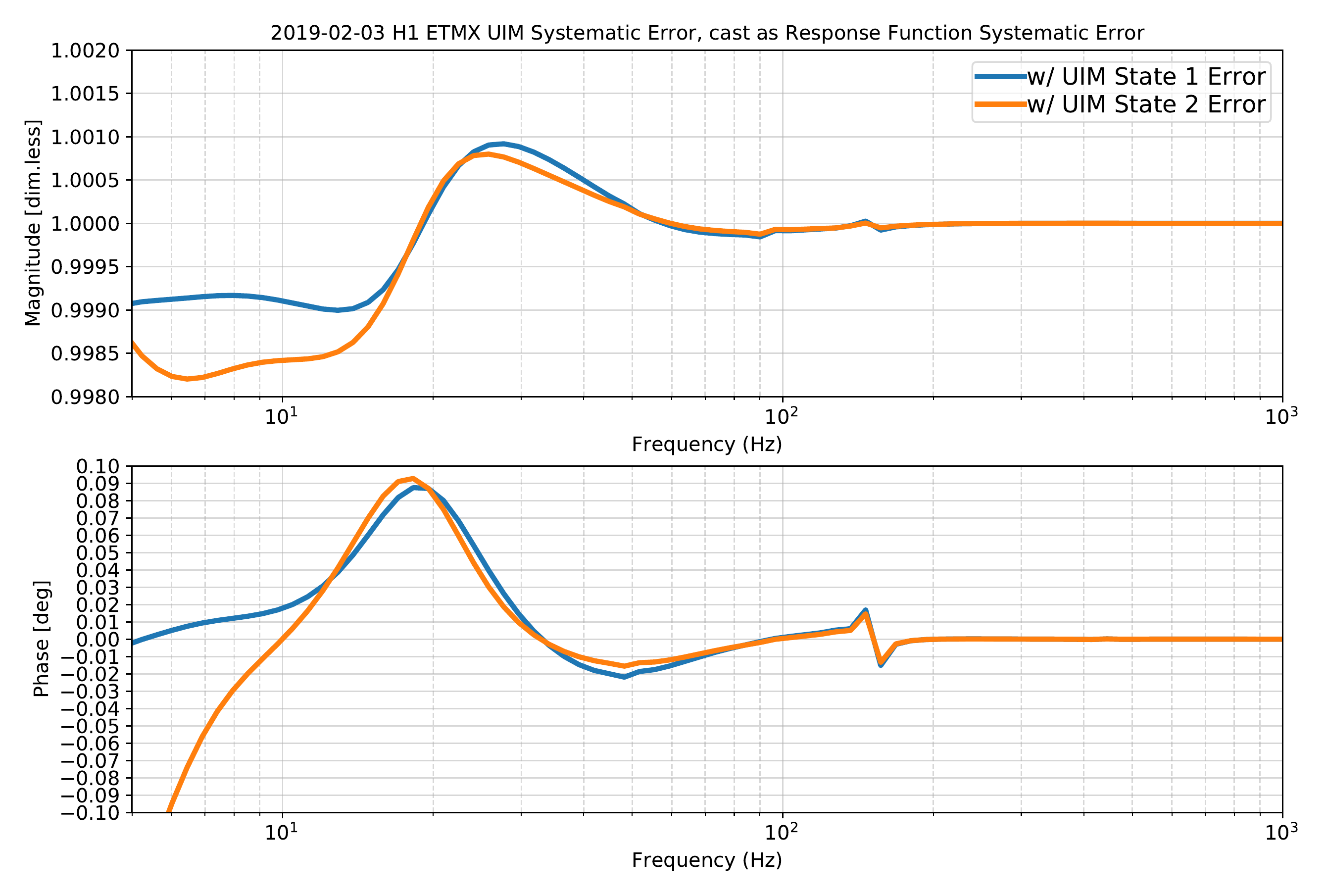}
	\caption[]{Systematic errors in the overall response function of the Hanford detector ($\eta_{R;A_{U}}$) due to the imperfect UIM electronics compensation ($\eta_{A_{U}}$). This is an unique issue at Hanford.}
	\label{fig:uimelectronicserror}
\end{figure}

For the vast majority of the entire O3 and during all measurements used to characterize the unknown systematic error in $A_{U}$, the UIM coil driver remained in State 1.
However, between November 27 and December 3, 2019, the configuration of the UIM coil driver was switched from State 1 to State 2, without knowing the accuracy of the compensation for State 2, and then switched back.
The accuracy of the compensation in both (a) State 1, used throughout most of O3, and (b) State 2, during this six-day period, was investigated in detail to quantify the resulting frequency-dependent error $\eta_{A_{U}}$; see details in \cite{Electronics-G2000527}. 
\Fref{fig:uimelectronicserror} shows the contribution of $\eta_{A_{U}}$ to the response function, i.e., $\eta_{R;A_{U}}$, for both configurations of the driver. 
Neither error in compensation, for State 1 or State 2, exceeds 0.15\% in magnitude and 0.09~deg in phase from 10~Hz to 1000~Hz, and is less than 0.05\% in magnitude and 0.03~deg in phase in the most sensitive region of the detector.
It is also well within the 68\% confidence interval of the GPR fit for unknown systematic error for the $A_{U}$ stage.
As such, this level of estimated contribution to the systematic error in the total response function, i.e., $\eta_{R;A_{U}}$, is deemed too small to be worth the effort to include in the collection of explicitly modeled systematic errors that contributes to $\eta_R$ in O3. 
However, because the level of impact was not identified and quantified until much after the most accurate, high-latency calibrated data stream was released, we excluded all measurements of $A_{U}$, $A_{P}$, and $A_{T}$ taken prior to December 4, 2019 from the collection of measurements used in the GPR to quantify unknown systematic error in each component of the actuation model. 
We acknowledge in retrospect that the exclusion of data is unnecessary, yet we believe there are sufficient measurements of each actuation stage taken between December 4, 2019 and the end of the run such that the GPR is sufficiently informed. And this particular $\eta_{R;A_{U}}$ discussed in this subsection is sufficiently covered in the resulting unknown systematic error estimate from the GPR fitting.

\subsubsection{TST resistance change in electrostatic drive signal path (O3B c boundary February 11, 2020)\\}
\label{sec:actuationelectronics-tst}

\begin{figure}[!tbh]
	\centering
	\includegraphics[width=0.8\textwidth]{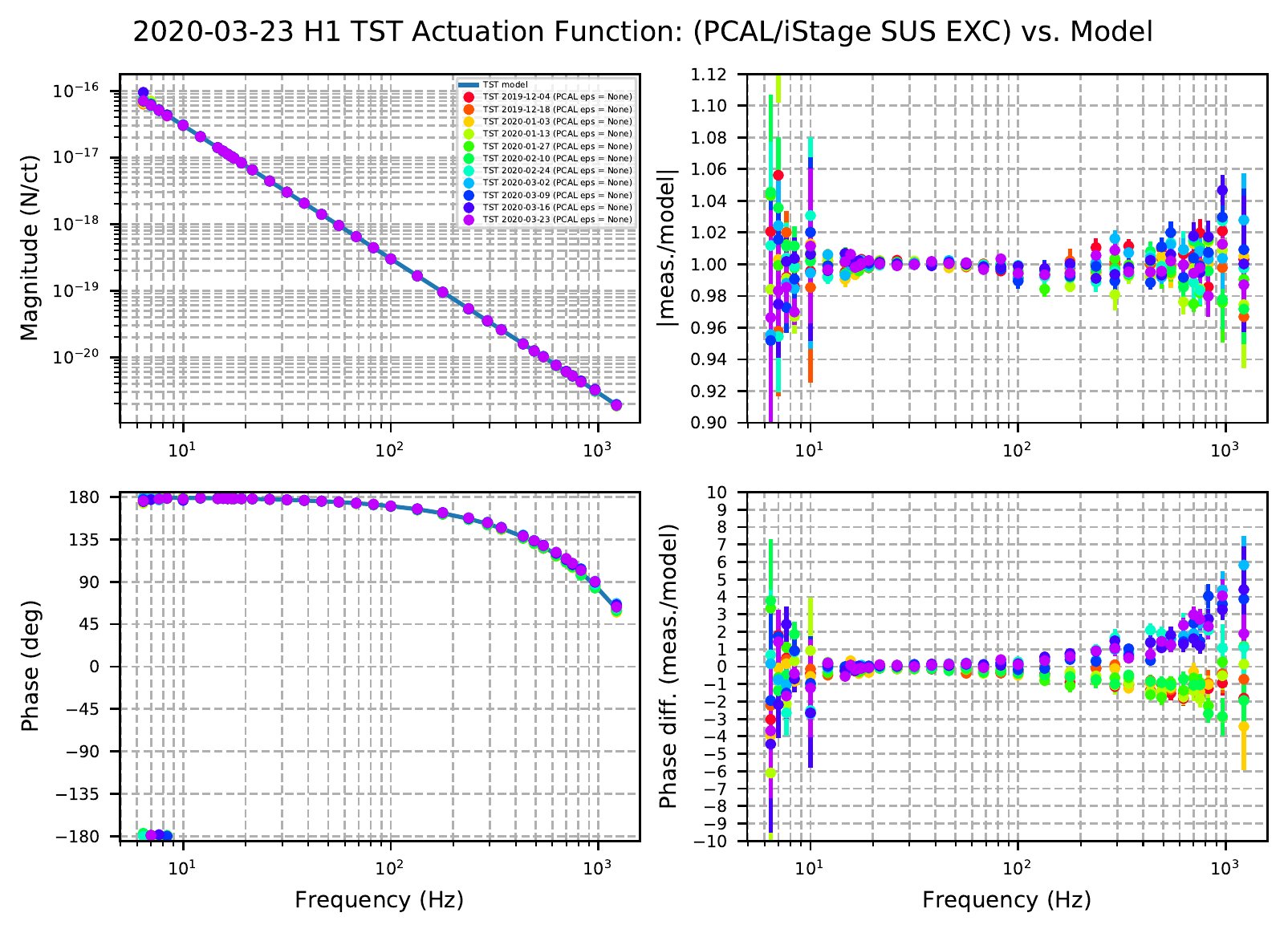}
	\caption[]{Comparison between the collection of TST-stage actuator measurements from December 4, 2019 through March 23, 2020 and the Hanford O3B TST-stage actuator model. The left panels are the magnitude and phase of the model and measured transfer functions, and the right panels are the magnitude and phase of the ratio, or ``residual'', with each measurement divided by the model.}
	\label{fig:tstinputtogpr}
\end{figure}

Changes to the TST electronics made on February 11, 2020, that were assumed to be inconsequential, did in fact cause some changes as revealed by the measurements made over the O3B interval.
The collection of TST actuation stage measurements, $A_{T}^{\rm {(meas)}}$ (Eq. (23) in~\cite{Sun2020}), taken from December 4, 2019 through March 23, 2020 compared against the model, $A_{T}^{\rm (model)}$ (Eq. (9) in~\cite{Sun2020}) is shown in \fref{fig:tstinputtogpr}.
While the magnitude residual is inconclusive, there is a clear bifurcation in the phase residual between February 10 and February 24, 2020.
The bifurcation in the data is consistent with a $\sim$4--5 deg phase change at 1~kHz. 

We suspect that this change is the result of the removal of a collection of current-limiting resistors (of resistance $\sim$10~k$\Omega$) from the signal path of the electrostatic drive (ESD) system.
These resistors were originally installed as a separate component to protect the sensitive electrodes when it became apparent that the high voltage ESD driver allowed for enough current to damage the in-vacuum electrodes.
The functionality of these independent resistors, however, was integrated into an upgrade of the ESD driver electronics before the start of O3. 
During the upgrade, the removal of the external redundant resistor collection was forgotten, leaving excess resistance in the signal path. 
Due to the large parasitic cable capacitance between the redundant resistors and the ESD electrodes (suspected to be of order $\sim$1~nF), the redundant 10-k$\Omega$ current-limiting resistors produced a voltage-divider-like, passive first-order low-pass filter response with a single pole frequency at
\begin{eqnarray}
\hspace{-1.0cm}f_{p}^{\textrm{LP}} & = & 1/(2\pi~[10^3 \textrm{Ohm}]~[10^{-9}~\textrm{Farad}]) = 15.9~\textrm{kHz},
\end{eqnarray}
throughout most of O3, prior to February 11, 2020. 
Such a low-pass filter incurs a phase loss, $\phi_{\textrm{LP}}$, at 1 kHz of 
\begin{eqnarray}
\hspace{-1.0cm}\phi_{\textrm{LP}} & = & (180/\pi) \tan^{-1}(-2 \pi [1000~\textrm{Hz}]~[10^3 \textrm{Ohm}]~[10^{-9}~\textrm{Farad}]) \nonumber\\ 
& = & -3.59~\textrm{deg}.
\end{eqnarray}
Upon the removal of these redundant resistors on February 11, 2020, the component was replaced by simple, so-called ``Safe High Voltage'' or SHV BNC barrel connectors with negligible resistance. 
We suspect that removing the phase loss from the unintended low-pass filter is the cause of the $\sim $4--5 deg phase increase at 1~kHz, as indicated by the residual $A_{T}^{\textrm{(meas)}}/A_{T}^{\textrm{(model)}}$ in \fref{fig:tstinputtogpr}. 

\begin{figure}[!tbh]
	\centering
	\includegraphics[width=0.8\textwidth]{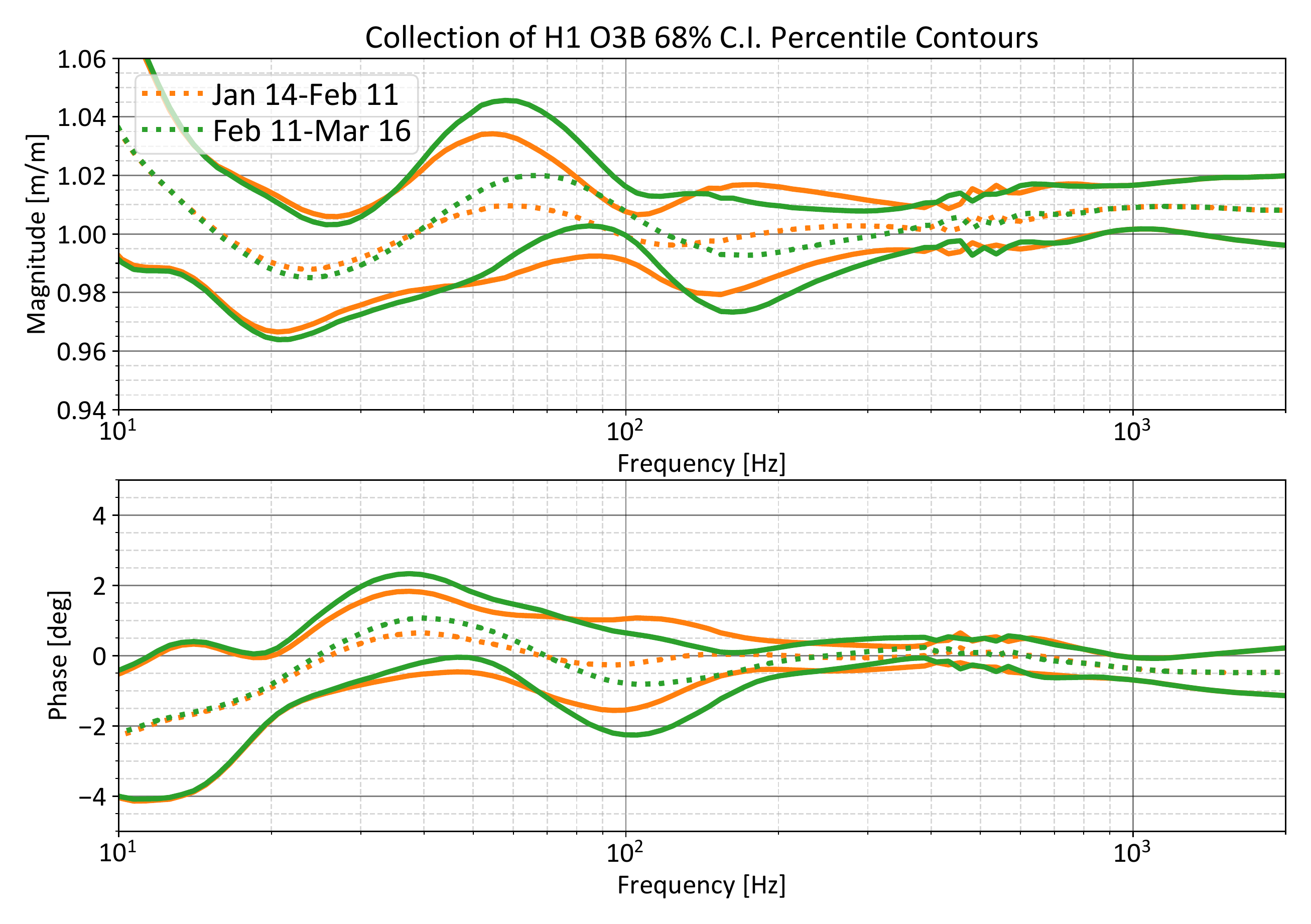}
	\caption[]{Change in the estimated systematic error in Hanford response function attributed to the change in electronics cabling in TST actuation function $A_{T}$ on February 11, 2020.}
	\label{fig:chunk2periodavsb}
\end{figure}

As we do not have a precise measurement of the parasitic cable capacitance, we cannot build a precise model of the low-pass filter, and thus cannot precisely account for the change in the actuation strength. 
As such, the collection of $A_{T}^{\textrm{(meas)}}$ measurements taken between December 4, 2019 and March 26, 2020 are divided into two distinct periods, with the boundary on February 11, 2020. 
We then use the standard GPR method to capture the ``unknown'' residual systematic error in the TST model ($\eta_{A_{T}}$) for each period separately, and propagate through the response function to quantify $\eta_{R;A_{T}}$ (see Eq. (11) in~\cite{Sun2020}).
\Fref{fig:chunk2periodavsb} shows the resulting impact on the overall systematic error estimate, comparing the estimate from January 14 to February 11, 2020 against that from February 11 to March 16, 2020.

\subsubsection{Changes of analog electronics in the sensing function\\}
\label{sec:sensingelectronics}

Similar to the driver electronics response in the actuation function, the sensing function also contains analog electronics with switchable frequency response. 
In this case, the gravitational-wave readout photodiodes (DCPDs) mounted on the output mode cleaner (OMC) have switchable levels of analog whitening.
As before, the response of the OMC DCPD whitening circuits are measured, fit, inverted, and installed in the digital signal path as compensation.
Unlike the UIM, the sensing function term heavily contributes to, if not dominates, the response function above 30~Hz, which requires more careful measurements and accurate compensation.
In fact, several poles in the whitening electronics analog response are compensated only with corrections added in the last stages of the acausal portion of the low-latency and high-latency calibration pipelines because they are above the Nyquist frequency of the digitally sampled data (see section 3.1.3 of~\cite{Sun2020}).
In the last week of O3B, on March 16, 2020, the configuration of the whitening electronics was switched from having all stages of its whitening response `ON' (State 3) to only having the first stage of whitening `ON' (State 1). 
Unfortunately, neither the circuit and the measurements thereof, nor the compensation were well-understood at the time. 
As a result, the low-frequency, sub-Nyquist compensation remained generally accurate, but the super-Nyquist poles that changed due to the configuration change were not updated and resulted in errors.

\begin{figure}[!tbh]
	\centering
	\includegraphics[width=\textwidth]{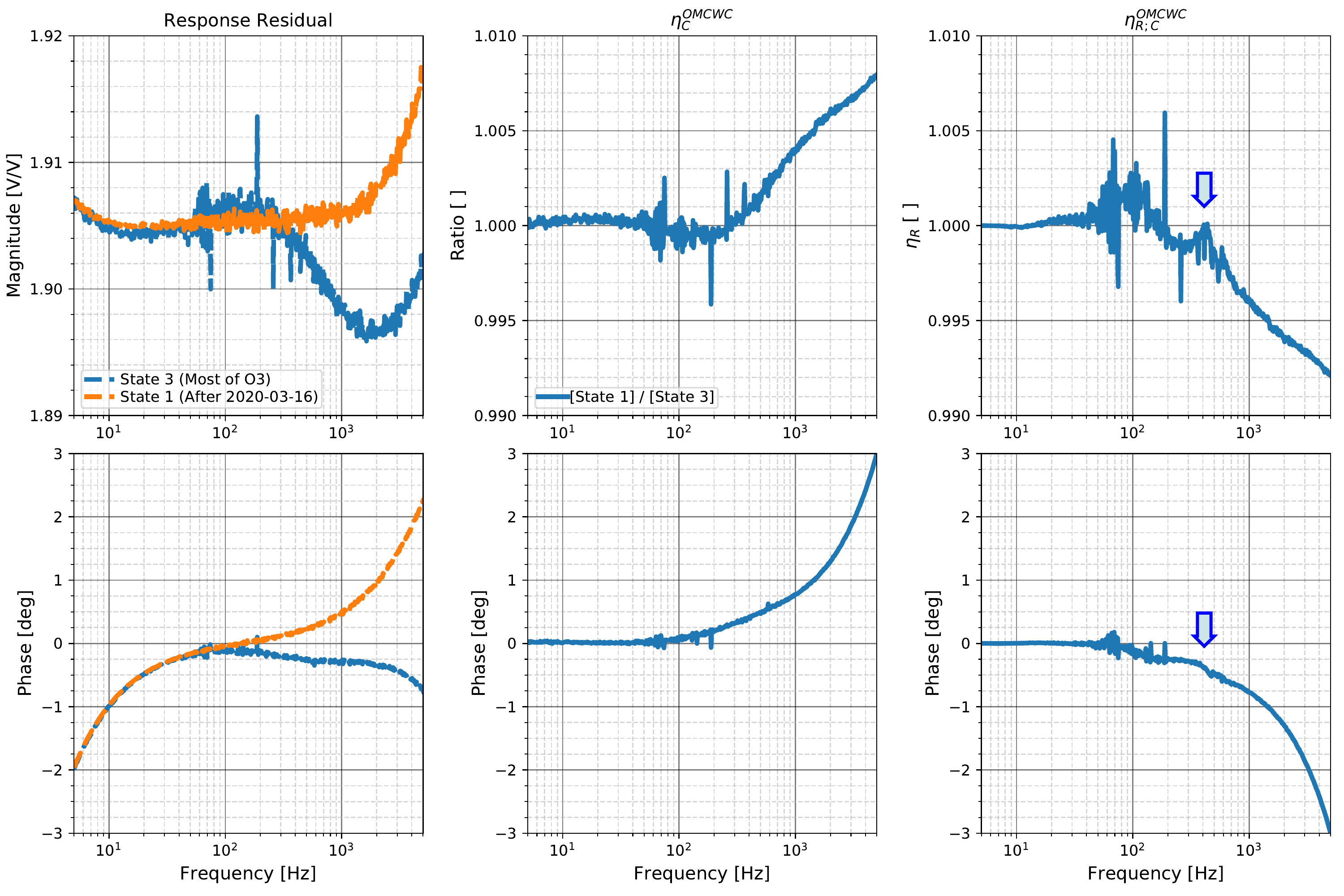}
	\caption[]{(Left) Bode plot of the transfer function of the measured analog whitening circuit response from the sum of the two DCPDs at Hanford, multiplied by their digital compensation filters and gain balancing coefficients.
		This is shown for ``State 3,'' in which all whitening stages are `ON', and ``State 1'' in which only the first whitening stage is `ON'. 
		(Middle) Bode plot of the ratio of the two transfer functions in the left panels. 
		This represents the systematic error, $\eta_{C}^{\rm OMCWC}$, incurred in the sensing function $C$, when the OMC whitening circuit (OMCWC) configuration is switched from State 3 to State 1 without compensating for the change in the super-Nyquist poles in the analog circuit response. 
		State 3 has more super-Nyquist poles than State 1, so the State 3 poles are over-compensating the number of poles in State 1, leading to the increase in the ratio of the transfer functions. 
		(Right) Bode plot of the systematic error in the detector response function, due to the error in the middle panels. 
		The calibration line frequency of 410.3 Hz is highlighted to show the apparent detector response function changes by 0.35 degrees at this frequency.}
	\label{fig:omcwcerror}
\end{figure}

\Fref{fig:omcwcerror} shows, in the left two panels, the sum transfer function of the two DCPDs that are used in the gravitational-wave readout, in each of the two configurations under question, including the digital compensation thereof. 
The middle two panels show the ratio of these two configurations transfer functions. 
This represents the error $\eta_{C}^{\rm OMCWC}$ incurred on the sensing function $C$, when the OMC whitening circuit (OMCWC) configuration is switched from State 3 to State 1 without updating the super-Nyquist poles in the response. 
Finally, the right two panels show the error in the response function $\eta_{R}$ caused by the $\eta_{C}^{\rm OMCWC}$. 
More details of the circuit, measurements, model, and fitting can be found in~\cite{Electronics-G2000527}.

The right most panels in \fref{fig:omcwcerror} highlight $f_{\rm Pcal} = 410.3$~Hz. 
This is one of the frequencies at which the Pcal system injects a constant, sinusoidal excitation on the detector as described in section 4.1 of~\cite{Sun2020}.
Recapping here, the excitation provides a reference for changes in the response function, cast as time-dependent correction factors (TDCFs) that are continuously applied to the data stream as in~\cite{Viets2018} (see the definitions and derivations of the TDCFs detailed in~\cite{TDCF-T1700106}).
The magnitude and phase of the response, however, are cast into corrections to the interferometric response of the sensing function: a change in the overall gain $\kappa_{C}$ and a change in the coupled cavity pole frequency $f_{cc}$,
\begin{eqnarray}
	\left(\frac{d_{err}(f_{\textrm{Pcal}})}{\Delta L_{\textrm{Pcal}}(f_{\textrm{Pcal}})}\right)\frac{1}{C_{R}} & = & S_{C}(f_{\rm Pcal},t) ~\equiv~ \frac{\kappa_{C}(t)}{1 + i f_{\textrm{Pcal}}/f_{cc}(t)},
\end{eqnarray}
which we may re-arrange and reduce to find
\begin{eqnarray}
	\kappa_{C} & = & |S_{C}(f_{\rm Pcal})| / \cos(\phi_{S_{C}}), \\
	f_{cc} & = & - f_{\textrm{Pcal}} \frac{\cos(\phi_{S_{C}})}{\sin(\phi_{S_{C}})},
\end{eqnarray} 
as in~\cite{TDCF-T1700106}.
The fixed change in switchable electronics response and the resulting systematic error from poor electronics compensation, however, does not depend on time. 
As such, the change in phase of $S_{C}$ at $f_{\textrm{Pcal}}$, $\delta$ (in units of deg), is misinterpreted by the TDCF system as a fixed bias $\Delta \kappa_{C}$ and $\Delta f_{cc}$ in $\kappa_{C}$ and $f_{cc}$, respectively, where
\begin{eqnarray}
	\Delta \kappa_{C} = \kappa_{C}' - \kappa_{C} & = & |S_{C}| \left[\frac{1}{\cos(\phi_{S_{C}} + \delta)} - \frac{1}{cos(\phi_{S_{C}})}\right], \\
	\Delta f_{cc} = f_{cc}' - f_{cc} & = & - f_{cal} \left[\frac{\cos(\phi_{S_{C}} + \delta)}{\sin(\phi_{S_{C}} + \delta)} - \frac{\cos(\phi_{S_{C}})}{\sin(\phi_{S_{C}})}\right].
\end{eqnarray}
We note explicitly that if $\delta = 0.35$~deg, the reported change $\Delta f_{cc}$ is 5~Hz, which impacts the response function in a broad frequency region down to $\sim$100~Hz.

These TDCFs $\kappa_{C}'$ and $f_{cc}'$, errantly biased by $\eta_{C}^{\rm OMCWC}$, have been applied to the data, producing an incorrect response function, immediately and continuously after the whitening circuit configuration change,
\begin{eqnarray}
	R_{\textrm{incorrect}} & = & \frac{1 + \kappa_{C}' \frac{1 + i f / f_{cc}^{\textrm{(model)}}}{1 + i f / f_{cc}'}~A^{\textrm{(model)}} ~D~C^{\textrm{(model)}}}{\kappa_{C}' \frac{1 + i f / f_{cc}^{\textrm{ref}}}{1 + i f / f_{cc}'}~C^{\textrm{(model)}}} \nonumber\\
	& = &  \frac{1 + \eta_{\textrm{TDCF}}^{ref \to inc}~A^{\textrm{(model)}} ~D~C^{\textrm{(model)}}}{\eta_{\textrm{TDCF}}^{ref \to inc}~C^{\textrm{(model)}}} \label{eq:Rincorrect},
\end{eqnarray}
where $\eta_{\textrm{TDCF}}^{ref \to inc}~\equiv~\kappa_{C}' \frac{1 + i f / f_{cc}^{\textrm{(model)}}}{1 + i f / f_{cc}'}$.

To create the correct response function, we must cancel out the applied errant TDCFs within $C'$, 
\begin{eqnarray}
R_{\textrm{correct}} & = & \frac{1 + \eta_{C}^{\rm OMCWC}~\left(\frac{\kappa_{C}}{\kappa_{C}'}~\frac{1 + i f / f_{cc}'}{1 + i f / f_{cc}}\right)~\eta_{\textrm{TDCF}}^{ref \to inc}~A^{\textrm{(model)}} D C^{\textrm{(model)}}}{\eta_{C}^{\rm OMCWC}~\left(\frac{\kappa_{C}}{\kappa_{C}'}~\frac{1 + i f / f_{cc}'}{1 + i f / f_{cc}}\right)~\eta_{\textrm{TDCF}}^{ref \to inc} C^{\textrm{(model)}}} \nonumber\\ 
& = & \frac{1 + \eta_{C}^{\rm OMCWC} \eta_{\textrm{TDCF}}^{inc \to cor} \eta_{\textrm{TDCF}}^{ref \to inc} A^{\textrm{(model)}}~D~C^{\textrm{(model)}}}{\eta_{C}^{\rm OMCWC}~\eta_{\textrm{TDCF}}^{inc \to cor} \eta_{\textrm{TDCF}}^{ref \to inc}~C^{\textrm{(model)}}} \label{eq:Rcorrect},
\end{eqnarray}
where $\eta_{\textrm{TDCF}}^{inc \to cor}~\equiv~\frac{\kappa_{C}}{\kappa_{C}'}~\frac{1 + i f / f_{cc}'}{1 + i f / f_{cc}}$.

From here, we can produce a time-independent systematic error in the response function due to the TDCF error as usual, denoted by $\eta_{R; \rm TCDF}$. 
The fundamental error due to the electronics change $\eta_{C}^{\rm OMCWC}$ has been modeled independently from measurements of the electronics (\fref{fig:omcwcerror}). 
The value of $A^{\textrm{(model)}}$ and $C^{\textrm{(model)}}$ are computed using the maximum a posteriori values of $\lambda_{\textrm{MAP}}^{A}$ and $\lambda_{\textrm{MAP}}^{C}$ as in Eq. (26) of~\cite{Sun2020} while $D$ is known completely.
We can obtain the values of ``before the change'' $\kappa_{C}$ and $f_{cc}$, and ``after the change'' $\kappa_{C}'$ and $f_{cc}'$ from what is calculated by the low-latency pipeline right before and right after the configuration change. 
The electronics configuration change can and has been made within 30 minutes while the detector is otherwise fully operational at high power and thermalized for many hours. 
As such, we can safely assume the actual (relative) optical gain and coupled cavity-pole frequency remain stable, and the only change in TDCF value is due to the circuit configuration change.
Putting it all together, using \eref{eq:Rincorrect} and \eref{eq:Rcorrect}, the total systematic error from this seemingly unimportant error in high-frequency electronics compensation is
\begin{eqnarray}
\hspace{-1.0cm}\eta_{R;\textrm{TDCF}} & = & \frac{R_{\textrm{correct}}}{R_{\textrm{incorrect}}} \nonumber\\
\hspace{-1.0cm} & = & \frac{1}{\eta_{C}^{\rm OMCWC}~\eta_{\textrm{TDCF}}^{inc \to cor}} \frac{1 + \eta_{C}^{\rm OMCWC} \eta_{\textrm{TDCF}}^{inc \to cor} \eta_{\textrm{TDCF}}^{ref \to inc} A^{\textrm{(model)}}~D~C^{\textrm{(model)}}}{\eta_{\textrm{TDCF}}^{ref \to inc}~A^{\textrm{(model)}}~D~C^{\textrm{(model)}}}. \nonumber\\
\end{eqnarray}
See~\cite{Electronics-G2000527} and~\cite{BadTDCFs-G2001293} for more details about the diagnosis of this problem.

\begin{figure}[!tbh]
	\centering
	\includegraphics[width=0.8\textwidth]{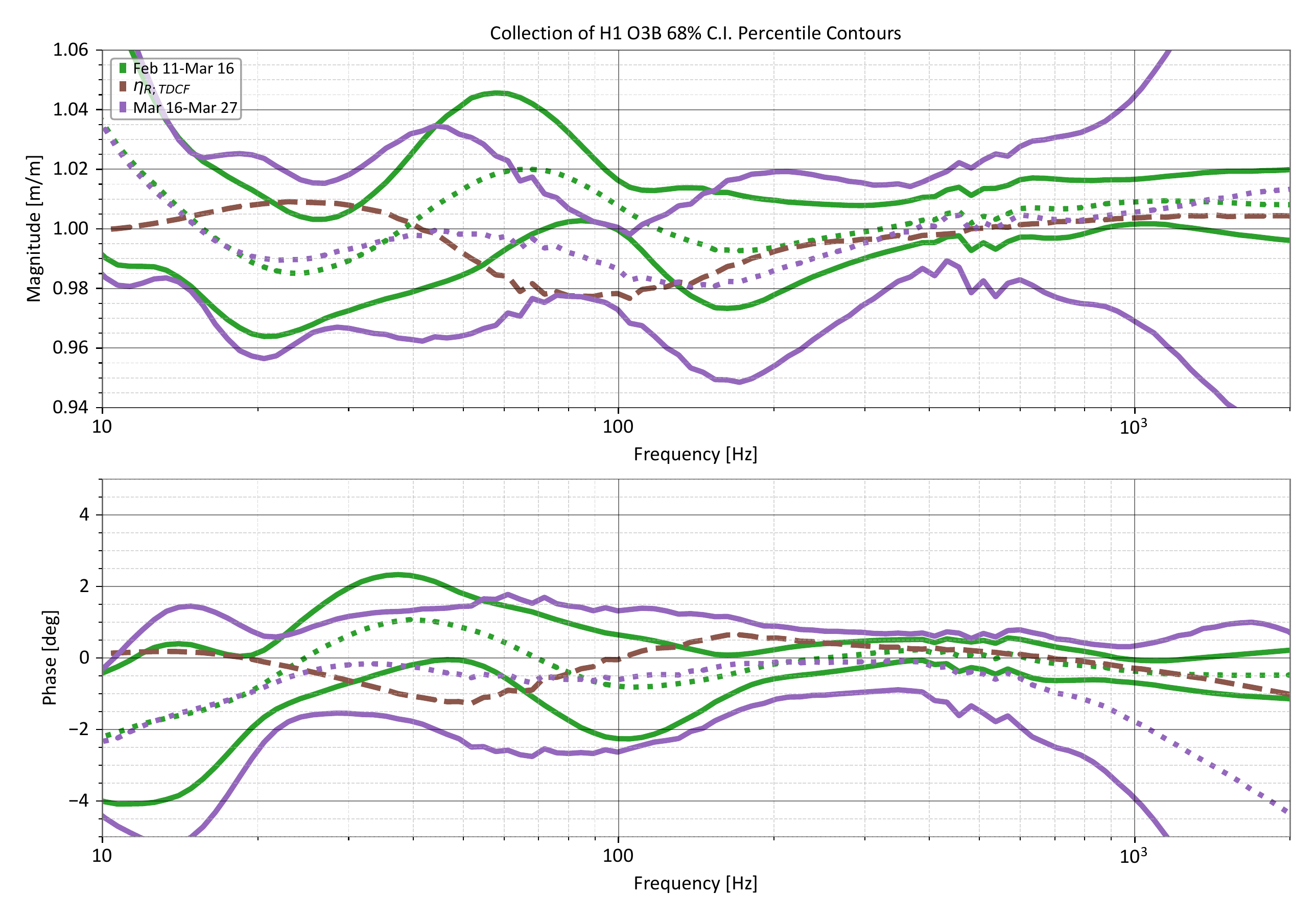}
	\caption[]{Change of the systematic error in the detector response function attributed to $\eta_{R;\textrm{TDCF}}$, due to the change in OMC whitening circuit configuration on March 16, 2020.}
	\label{fig:chunk2periodbvsc}
\end{figure}

\Fref{fig:chunk2periodbvsc} shows the resulting systematic error $\eta_{R;\textrm{TDCF}}$ (brown dashed curve). This systematic error is deemed large enough to be included as a multiplicative term in all the hourly estimates of the overall systematic error and statistical uncertainty in the detector response function from March 16 to March 27, 2020. 
\Fref{fig:chunk2periodbvsc} also shows the statistical summary of the systematic error estimates in the detector response function (median and 1$\sigma$ uncertainty boundaries), collected from all hourly estimates in the two different periods.
While the dominant change from the period February 11--March 16, 2020 to March 16--March 27, 2020 is from $\eta_{R;\textrm{TDCF}}$, the unknown systematic error in the latter period is also re-assessed with GPR using only the single measurement taken on March 23, 2020, and a small collection of the long-duration discrete sinusoidal excitations above 1 kHz captured during that period. 
With so few measurements between March 16 and March 27, 2020 to inform the residual error and statistical uncertainty using GPR, the 1$\sigma$ uncertainty in the sensing function error estimate increases dramatically beyond prior periods. 
In hindsight, this reduction in the collection of data used as input to the GPR fit for unknown systematic error is unnecessary, since the model for $\eta_{R;\textrm{TDCF}}$ is well-quantified and precise, i.e., we could have used the entire O3B collection of $C^{(meas)}$ in the GPR fit and multiplied by $\eta_{R;\textrm{TDCF}}$. 
In \fref{fig:chunk2periodbvsc}, this would be the equivalent to multiplying the green curves by the brown dashed curve. 
Instead, we have the purple curves, which accurately reflects the change, but likely over-estimates the 68\% confidence interval above $\sim$100~Hz. 

\section{Livingston detector changes}
\label{sec:llo}
During O3B, several changes directly impact the DARM control loop at the Livingston detector and thus require changes to the calibration model used to calibrate the strain data.
These changes are: 1) a 30\% increase in the digital gain of the Y-arm end test mass drive strength to compensate for a slowly physically weakening ESD; 2) a change of the X-arm coil driver electronics in the path to drive the penultimate mass (PUM) of the quadruple test mass suspension; and 3) a switch from sending the DARM drive control from the Y-arm to the X-arm end test mass.
In each case, the effects of these changes were measured and accounted for in the final control loop model used to calibrate the strain data.

\subsection{X-arm electrostatic drive gain}
\label{sec:lloESDgain}
The strength of the ESDs is known to drift over time, and is generally associated with charge accumulation on the test mass~\cite{buikema2020sensitivity,prokhorov2010space}.
During O3B, the Y-arm ESD actuator had become 30\% weaker compared to the start of O3.
This leads to several undesirable effects including reducing the unity gain frequency of the DARM control loop, which affects interferometer stability, and increasing the systematic error and uncertainty estimate in the detector response in a band of 40--60 Hz due to smaller actuation strength when the test mass stage and the penultimate mass stage actuation are combined.

\begin{figure}[!tbh]
        \centering
        \subfigure[]
        {
                \label{fig:before_contri}
                \scalebox{0.3}{\includegraphics{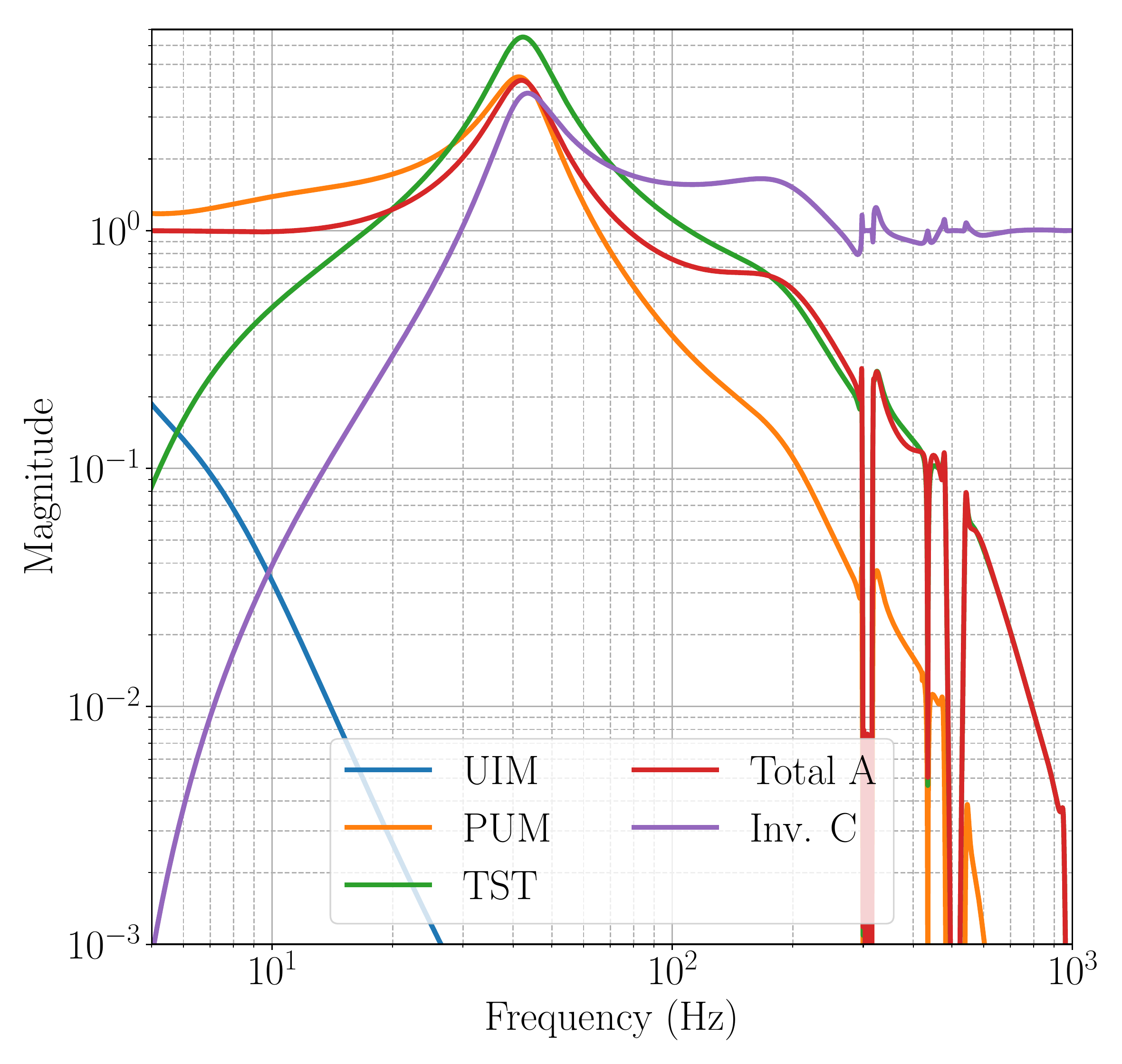}}
        }
        \subfigure[]
        {
                \label{fig:after_contri}
                \scalebox{0.3}{\includegraphics{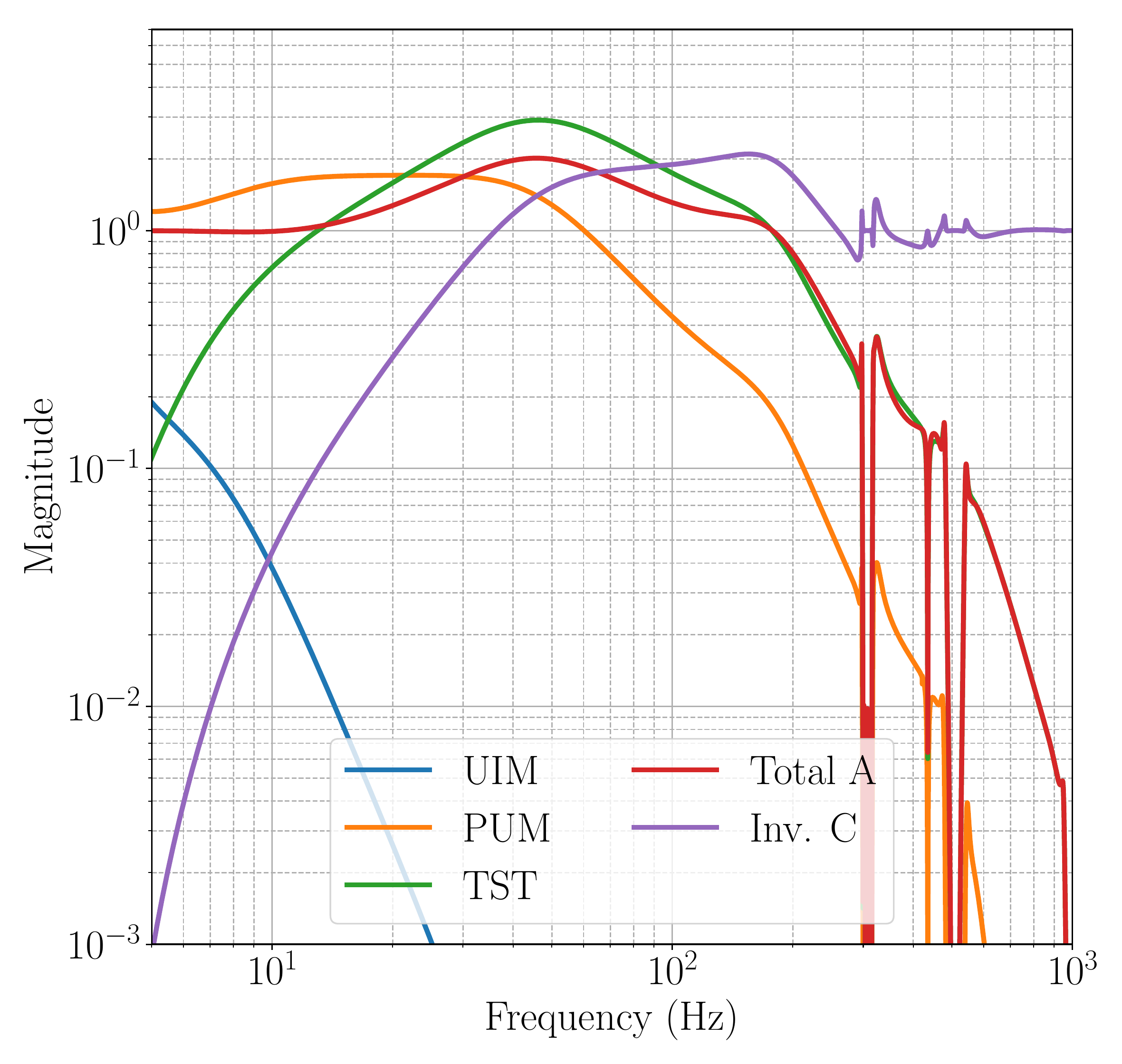}}
        }
        \caption[]{At Livingston the contributions of each stage of the bottom three end test mass actuators, namely the Y-arm test mass (TST), X-arm penultimate test mass (PUM), and X-arm upper intermediate mass (UIM), as well as the overall actuation (A), and the inverse sensing (Inv. C) to the response function (a) before and (b) after the 30\% increase in the TST digital drive strength on January 14th, 2020.}
        \label{fig:contri}
\end{figure}

On January 14, 2020, the digital drive gain to the Y-arm ESD was increased by 30\% to compensate for this observed overall drift.
In \fref{fig:contri}, we show the contribution of each actuation stage, total actuation function, and sensing function to the overall response function before and after the drive gain change.
This change improves the overall combined systematic error and uncertainty by a factor of $\sim3$ in the 40--50~Hz band due to how the test mass stage and penultimate stage actuations are combined in the total actuation.

\subsection{Penultimate mass coil driver electronics change}
\label{sec:lloPUM}
On February 11, 2020, the electronics that drive the coils used to actuate the X-arm PUM stage of the quadruple suspension were exchanged for a newer version, with a slightly different transfer function.
However, the interferometer digital compensation filters were not updated to use the measured transfer function for these new electronics until February 25, 2020.
To account for this discrepancy, the calibration model used for the most accurate high-latency version of strain data was adjusted to take into account the very slight difference ($<1\%$ in magnitude) between the new analog electronics transfer function and the outdated digital compensation.

\begin{figure}
        \begin{center}
                \includegraphics[width=0.8\textwidth]{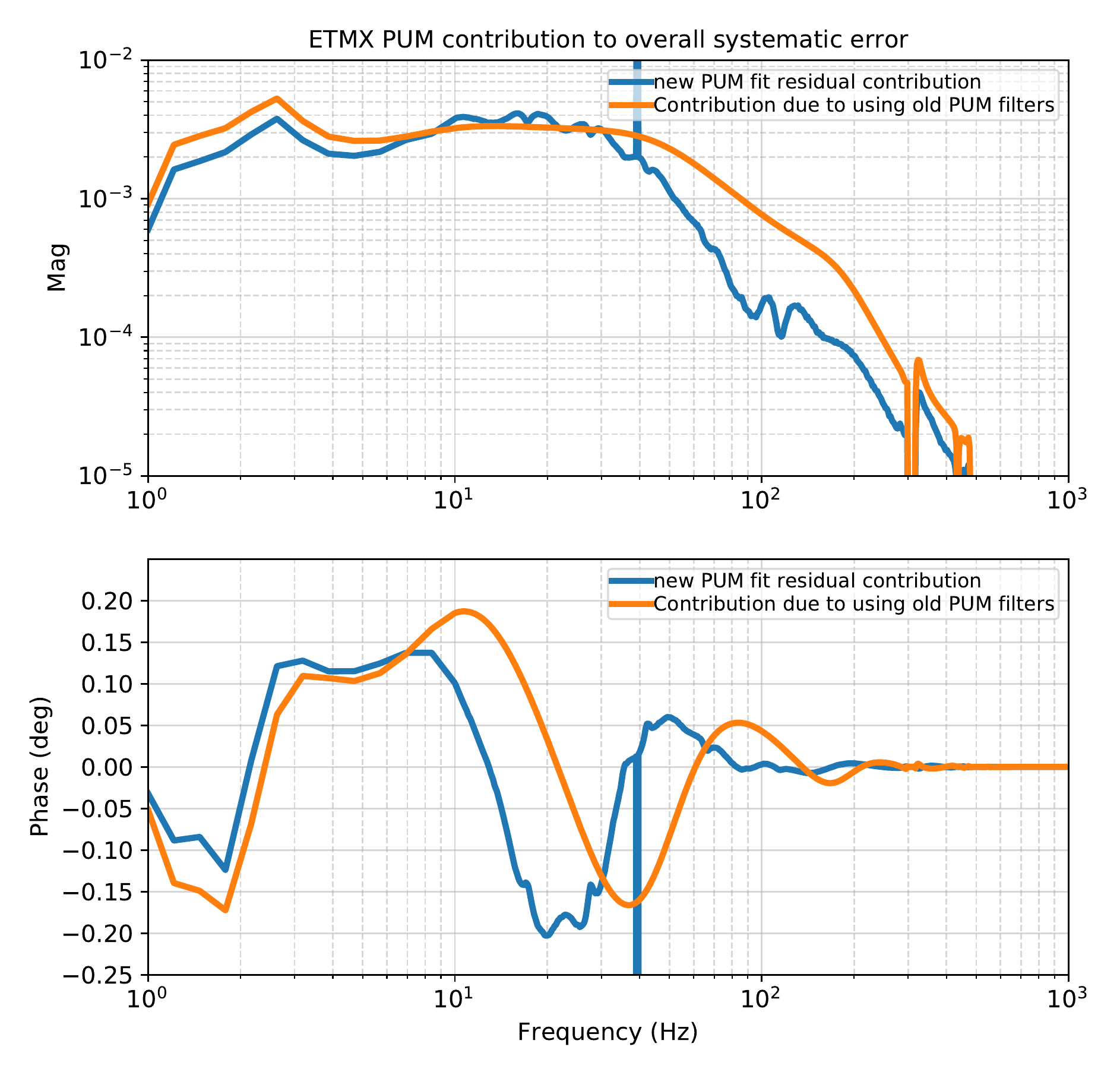}
        \end{center}
        \caption{At Livingston the error in the overall response function due to the error in the PUM coil driver electronics alone. The blue curve shows the expected contribution to the overall systematic error due to the residual of the fits to the new PUM coil driver electronics. The orange curve shows the expected contribution to overall systematic error when using the old digital compensation with the new analog electronics, assuming that the fits for both new and old electronics are perfect.
        \label{fig:pum_residual}}
\end{figure}

During this two-week time period, the low-latency calibrated strain data was slightly affected, with a small error $<1\%$ in magnitude and $< 0.25$~deg in phase.
\Fref{fig:pum_residual} shows the estimated contribution to the overall systematic error in calibrated data due to the residual error between the measurements and the electronics fits used to generate the new digital compensation filters.
It also shows the estimated contribution to the overall systematic error introduced in the low-latency calibrated strain data (but not the final high-latency calibrated strain data) by using the outdated digital filters with the new electronics.

We note that this is only one component of the actuation transfer function for the PUM stage.
GPR analyses of multiple transfer function measurements performed every one to two weeks during the run allow further quantifying the systematic error and uncertainty in the whole PUM stage~\cite{Sun2020}.

\subsection{DARM test mass actuation switched from Y arm to X arm}
\label{sec:lloDARMswitch}
For all of O3A and the majority of O3B, the DARM control signal at the Livingston detector was sent to the TST stage of the Y-arm end suspension and the two middle stages (UIM and PUM) of the X-arm end suspension.
In an effort to explore possible noise improvements to the interferometer, this control scheme was changed to send the DARM control signal to the TST stage of the X-arm end suspension on March 10, 2020.

\begin{figure}
        \begin{center}
                \includegraphics[width=0.8\textwidth]{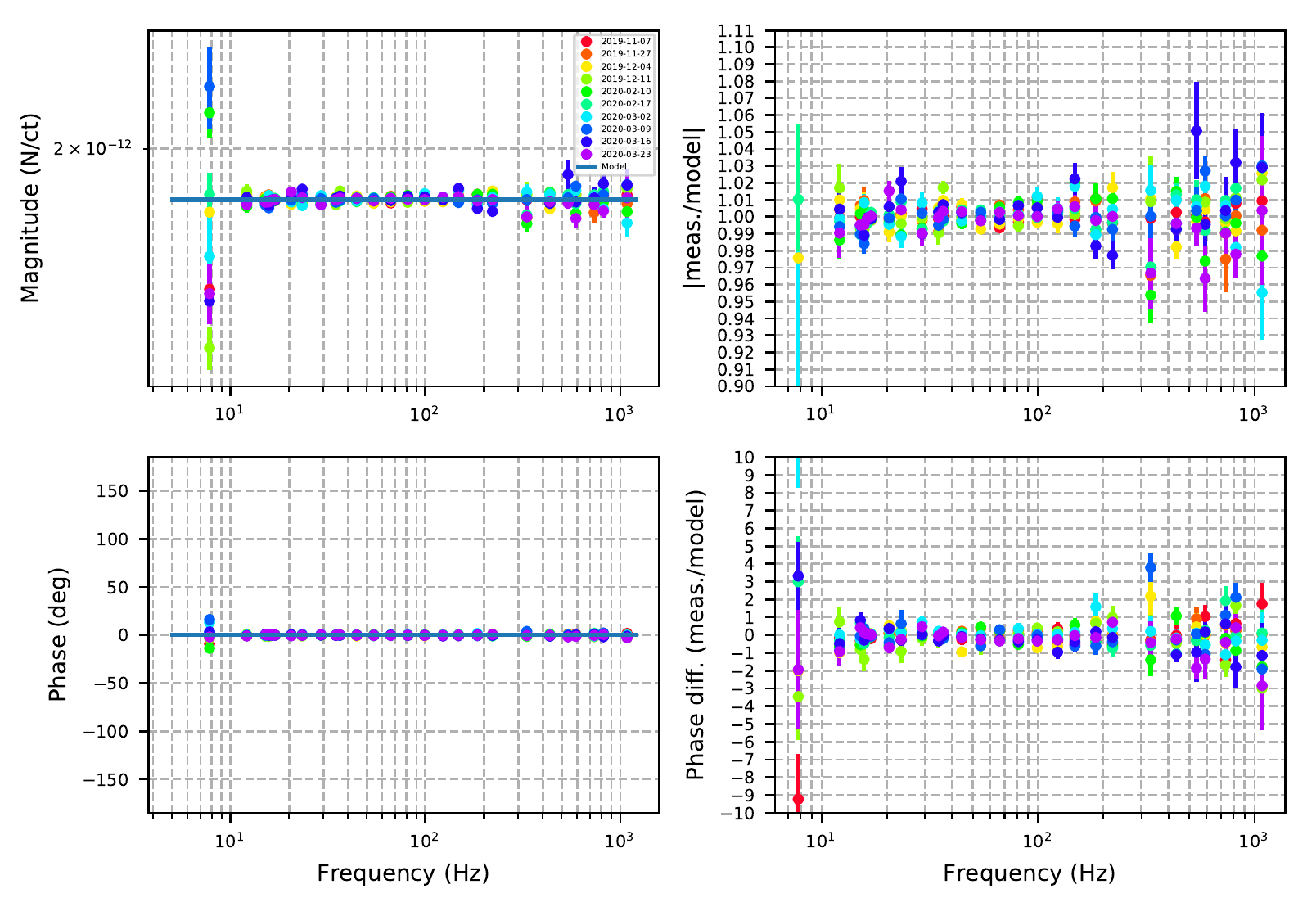}
        \end{center}
        \caption{Transfer function measurements at Livingston of the X-arm and Y-arm TST actuators taken throughout O3B (left) and the fractional residuals between the measurements and the corresponding calibration model (right).  Measurements prior to March 10, 2020 use the Y-arm TST actuator, and the two after use the X-arm TST actuator. 
        \label{fig:y_x_sweeps}}
\end{figure}

The net effect of this change on the calibration is only a frequency-independent gain in terms of the physical actuation strength of the TST stage, which has been compensated for with digital gains and tracked by calibration lines.
The digital compensation and modeling of the actuation of these two test masses sufficiently accounts for this change such that the transfer function measurements show no significant discrepancy in the frequency bands where the TST actuator dominantly contributes to the overall response function.
In \fref{fig:y_x_sweeps}, we plot these corrected measurements on top of each other for comparison.

In practice, we switched the TST calibration line driven on the TST stage of the Y-arm end suspension to the equivalent X-arm end suspension, requiring a small update to the calibration pipeline to accommodate the new control scheme.
Overall, the change was transparent to end users of the calibrated strain data.

\section{GPR parameter choices}
\label{sec:gpr}






Further investigations of the GPR fitting parameters used to infer unknown systematic errors in $A_{i}$ and $C$ motivated the changes listed in \tref{tab:GPRval} for O3B.
An introduction and review of the GPR process in general can be found in~\cite{GPR}.
Of interest here, while the kernel remains the same as O3A (c.f. (24) and surrounding text in~\cite{Sun2020}), comprised of a modified Gaussian-type radial basis function (RBF, \cite{buhmann2003radial}), the values for the length scale hyperparameters, $\ell$, and the frequency regions of data being used to inform the regressions have been updated.
These parameter investigation results are summarized here, with the full details provided in~\cite{LHOaLOG55012}.

First, for the Hanford sensing function, the lower frequency limit, $f_{\textrm{min}}$, of the $C^{\textrm{(meas)}}$ data used to inform the regression is adjusted upwards from 12~Hz to 20~Hz because the region below 20~Hz is dominated by deficiencies in the sensing function model (c.f. section 4.2 of~\cite{Sun2020} and \sref{sec:thermal} in this article). 
This change in $f_{\textrm{min}}$ increases the 68\% confidence interval of the regression at frequencies $\lesssim 20$~Hz to a level deemed appropriate to reflect that deficiency.
This increase in the 68\% confidence interval at lower frequencies may result in an overestimate of $\eta_{R}$ due to the slightly overestimated sensing function error, $\eta_{C}$.
The collection of sensing function measurements $C^{\textrm{(meas)}}$, used to inform the GPR, do not account for thermalization effects (as described in \sref{sec:thermal}).
Some thermalization effects are already present in $\eta_{C}$ as the GPR output using $C^{\textrm{(meas)}}$, and we further account for thermalization effects at the beginning of each observing interval when the Hanford detector has just reached its low-noise operating configuration.
Thus the low frequency contribution to $\eta_{R}$ from $\eta_{C}$ is slightly overestimated for some hourly estimates.
Since the Livingston detector does not suffer from any of these deficiencies, the lower limit remains unchanged from O3A.

Second, the $f_\textrm{min}$ values for the collection of $A_{i}^{\textrm{(meas)}}$ to inform the regressions of actuation functions are adjusted upwards from 6~Hz to 7~Hz for $A_{U}$ and $A_{P}$, and to 10~Hz for $A_{T}$ at both detectors. 
The low-frequency noise below 10~Hz tends to be highly non-stationary and difficult to measure due to the rapidly decreasing detector sensitivity in that frequency range.
The higher strength of the coil actuators involved in $A_{U}$ and $A_{P}$ allows for reproducible results to slightly lower frequencies.

Third, the upper frequency limits, $f_\textrm{max}$, of the $A_{i}^{\textrm{(meas)}}$ data used to inform the regression are increased for the UIM and PUM stages (see \tref{tab:GPRval}). Modeling of the narrow-band, high-quality-factor features in the quadruple pendulum dynamics has been improved in O3B, reducing the sharp residual systematic error at $\sim 10^2$~Hz due to the model deficiencies that previously impact GPR.
As such, more data at higher frequencies can be used in the fit with the updated length scales.

Finally, the RBF length scale, $\ell$, of the GPR was modified for each of the sensing and actuation functions in order to plausibly account for or identify any previously unknown systematic errors.
In this application of the RBF kernel in the GPR, the length scale is related to frequency through $df = \pm f (10^{\pm \ell} - 1)$~\cite{GPR-length-scale}.
Thus, narrow-band deviations or systematic errors are characterized by small values of $\ell$ whereas broad-band errors would be characterized by large $\ell$ values.
Changes to the prior of $\ell$ allow the GPR to better identify such errors if they are present in the data.
No significant unknown systematic errors were found in O3B because of these changes.

\begin{table}[!tbh]
	\centering
	\caption{\label{tab:GPRval}Parameter limits and frequency intervals used for Gaussian Process Regression in O3A and O3B.}
		\setlength{\tabcolsep}{5pt}
		\begin{tabular}{llll}
		\br
		Detector & Parameter & O3A value & O3B value \\
		\mr
		H1 & Sensing [$f_\textrm{min}$, $f_\textrm{max}$] (Hz) & [12, 5000] & [20, 5000] \\
		H1 & Sensing RBF $\ell$ limits [min, max] & [0.48, 0.52] & [0.33, 0.37] \\
		L1 & Sensing [$f_\textrm{min}$, $f_\textrm{max}$] (Hz) & [12, 5000] & [12, 5000] \\
		L1 & Sensing RBF $\ell$ limits [min, max] & [0.48, 0.52] & [0.33, 0.37] \\
		H1/L1 & UIM [$f_\textrm{min}$, $f_\textrm{max}$] (Hz) & [6, 50] & [7, 250] \\
		H1/L1 & UIM RBF $\ell$ limits [min, max] & [0.5, 1.5] & [0.08, 0.12] \\
		H1/L1 & PUM [$f_\textrm{min}$, $f_\textrm{max}$] (Hz) & [6, 400] & [7, 500] \\
		H1/L1 & PUM RBF $\ell$ limits [min, max] & [0.1, 0.5] & [0.18, 0.22] \\
		H1/L1 & TST [$f_\textrm{min}$, $f_\textrm{max}$] (Hz) & [6, 1000] & [10, 1000] \\
		H1/L1 & TST RBF $\ell$ limits [min, max] & [0.1, 0.5] & [0.48, 0.52] \\ 
		\br
		\end{tabular}
\end{table}

\section{Conclusion}
\label{sec:conclusion}
In this paper, we presented the systematic error and statistical uncertainty in the most accurate, high-latency version of strain data  used for gravitational-wave astrophysical parameter estimation in O3B at the Hanford and Livingston detectors.
We discuss the improvements in understanding the detector systematic errors and highlight challenges specific to the O3B portion of the observing run, propagating their impact through to the estimated systematic error in the overall detector response function.
We find the levels of the overall systematic error roughly consistent with O3A, and yet in each epoch, the changes of the detector configuration have introduced non-negligible impact on the frequency-dependent systematic error, leading to larger uncertainties and/or larger errors during some observational time.
In O3B, the overall, combined systematic error and associated uncertainty of the most accurate calibrated data is within $\sim 10$\% in magnitude and 10 deg in phase in the frequency band 20–2000 Hz. In this same band, the systematic error alone is estimated to be below 2\% in magnitude and 5 deg in phase. 

The understanding of the Pcal systems and thus the uncertainty and systematic error of this absolute reference continue to improve in O3B. 
However, we show that the combined systematic error and uncertainty in the detector response function are dominated by errors and uncertainties in many individual frequency-dependent terms in the response function model rather than in the absolute reference at this stage.  

It remains true that the detection of transient gravitational waves is insensitive to the level of systematic error and uncertainty now regularly achieved~\cite{Lindblom2009, abbott2016gw150914}.
However, in the face of this practical limit of systematic errors in calibration, understanding of the effects on uncertainty and bias in astrophysical parameter estimation continues to improve as studies are carried out to better integrate these errors. 
For example, in Refs.~\cite{vitale2020physical} and \cite{payne2020gravitational}, the authors have used results from the first gravitational wave transient catalog (GWTC-1 \cite{GWTC-1}) and the systematic error estimated in Advanced LIGO's first and second observing runs (O1 and O2)~\cite{cahillane2017calibration}, and concluded that the astrophysical parameter estimation for the GWTC-1 events is insensitive to calibration error at the levels described in~\cite{cahillane2017calibration}.
However, in O1 and O2, the durations of the observation were short (with much less error-prone configuration changes and parameter drift), the detectors were operating at low power (with much less deficiencies in modeling the sensing function), and our methods for characterizing the errors and the understanding of the details were not as sophisticated as those achieved in O3~\cite{Sun2020}.
In addition, the GWTC-1 events have relatively low signal-to-noise ratio given that the detectors were less sensitive in O1 and O2.
These facts imply that the combined effects of calibration errors, event signal-to-noise ratios, and accuracy and precision of astrophysical parameter estimation have not yet been sufficiently explored. 
Quantitative requirements on the level of calibration errors and uncertainties for future detectors therefore remain to be studied. 
As such, investigations of the impacts on transient gravitational wave events at high SNRs, using the realistic systematic errors identified in O3 as examples for future observations, are under way. 
Beyond the impact on parameter estimation for individual events, studies are also underway to investigate the impact on cosmology and tests of General Relativity using collections of events detected in different calibration epochs, e.g., bias in the estimates of the Hubble constant or limits on statistical averaging.

\section{Acknowledgments}
The authors gratefully acknowledge the operators, commissioners, and LSC fellows at Hanford and Livingston for their help in setting up the detector configurations and taking measurements needed for this work.
This material is based upon work supported by NSF's LIGO Laboratory which is a major facility fully funded by the National Science Foundation.
LIGO was constructed by the California Institute of Technology and Massachusetts Institute of Technology with funding from the NSF, and operates under cooperative agreement PHY--1764464. Advanced LIGO was built under award PHY--0823459.
The authors gratefully acknowledge the support of the United States NSF for the construction and operation of the
LIGO Laboratory and Advanced LIGO as well as the Science and Technology Facilities Council (STFC) of the
United Kingdom, the Max-Planck-Society (MPS), and the State of
Niedersachsen/Germany for support of the construction of Advanced LIGO
and construction and operation of the GEO600 detector.
Additional support for Advanced LIGO was provided by the Australian Research Council (ARC).
LS, DB, PBC, LEHD, TM, EP and CC acknowledge the LSC Fellows program for supporting their research at LIGO sites.
EG acknowledges the support of the Natural Sciences and Engineering Research Council (NSERC) of Canada.
DB and SK are supported by NSF award PHY--1921006.
AV is supported by NSF award PHY--1841480.
MW is supported by NSF awards PHY--1607178 and PHY--1847350.
LS and EP acknowledge the supported of the ARC Centre of Excellence for Gravitational Wave Discovery (OzGrav), grant number CE170100004.
PBC acknowledges the support of the Spanish Agencia Estatal de Investigaci{\'o}n and Ministerio de Ciencia, Innovaci{\'o}n y Universidades grants FPA2016-76821-P, the Vicepresidencia i Conselleria d'Innovaci{\'o}, Recerca i Turisme del Govern de les Illes Balears (grant FPI-CAIB FPI/2134/2018), the Fons Social Europeu 2014-2020 de les Illes Balears, the European Union FEDER funds, and the EU COST actions CA16104, CA16214, CA17137 and CA18108.
The authors would like to thank all of the essential workers who put their health at risk during the COVID-19 pandemic,
without whom we would not have been able to complete this work.
This paper carries LIGO Document Number \dcc.

\clearpage
\newpage
\providecommand{\newblock}{}


\begin{thebibliography}{10}
	\expandafter\ifx\csname url\endcsname\relax
	\def\url#1{{\tt #1}}\fi
	\expandafter\ifx\csname urlprefix\endcsname\relax\def\urlprefix{URL }\fi
	\providecommand{\eprint}[2][]{\url{#2}}
	
	\bibitem{Sun2020}
	Sun L, Goetz E, Kissel J~S, Betzwieser J, Karki S, Viets A, Wade M,
	Bhattacharjee D, Bossilkov V, Covas P~B, Datrier L~E~H, Gray R, Kandhasamy S,
	Lecoeuche Y~K, Mendell G, Mistry T, Payne E, Savage R~L, Weinstein A~J, Aston
	S, Buikema A, Cahillane C, Driggers J~C, Dwyer S~E, Kumar R and Urban A 2020
	{\em Classical and Quantum Gravity\/} {\bf 37} 225008
	\urlprefix\url{https://doi.org/10.1088/1361-6382/abb14e}
	
	\bibitem{LIGO2014}
	Aasi J {\em et~al.\/} (LSC) 2015 {\em Classical and Quantum Gravity\/} {\bf 32}
	074001 \urlprefix\url{https://doi.org/10.1088/0264-9381/32/7/074001}
	
	\bibitem{Virgo2014}
	Acernese F {\em et~al.\/} (Virgo) 2015 {\em Classical and Quantum Gravity\/}
	{\bf 32} 024001 \urlprefix\url{https://doi.org/10.1088/0264-9381/32/2/024001}
	
	\bibitem{GWTC-1}
	Abbott B~P {\em et~al.\/} (LIGO Scientific Collaboration and Virgo
	Collaboration) 2019 {\em Phys. Rev. X\/} {\bf 9}(3) 031040
	\urlprefix\url{https://link.aps.org/doi/10.1103/PhysRevX.9.031040}
	
	\bibitem{GWTC-2}
	Abbott R, Abbott T, Abraham S, Acernese F, Ackley K, Adams A, Adams C, Adhikari
	R, Adya V, Affeldt C {\em et~al.\/} 2020 {\em arXiv preprint
		arXiv:2010.14527\/}
	
	\bibitem{buikema2020sensitivity}
	Buikema A, Cahillane C, Mansell G, Blair C, Abbott R, Adams C, Adhikari R,
	Ananyeva A, Appert S, Arai K {\em et~al.\/} 2020 {\em Physical Review D\/}
	{\bf 102} 062003
	
	\bibitem{Cahillane2017}
	Cahillane C, Betzwieser J, Brown D~A, Goetz E, Hall E~D, Izumi K, Kandhasamy S,
	Karki S, Kissel J~S, Mendell G, Savage R~L, Tuyenbayev D, Urban A, Viets A,
	Wade M and Weinstein A~J 2017 {\em Phys. Rev. D\/} {\bf 96}(10) 102001
	\urlprefix\url{https://link.aps.org/doi/10.1103/PhysRevD.96.102001}
	
	\bibitem{rollins2016distributed}
	Rollins J~G 2016 {\em Review of Scientific Instruments\/} {\bf 87} 094502
	
	\bibitem{Karki2016}
	Karki S, Tuyenbayev D, Kandhasamy S, Abbott B~P, Abbott T~D, Anders E~H,
	Berliner J, Betzwieser J, Cahillane C, Canete L, Conley C, Daveloza H~P,
	De~Lillo N, Gleason J~R, Goetz E, Izumi K, Kissel J~S, Mendell G, Quetschke
	V, Rodruck M, Sachdev S, Sadecki T, Schwinberg P~B, Sottile A, Wade M,
	Weinstein A~J, West M and Savage R~L 2016 {\em Review of Scientific
		Instruments\/} {\bf 87} 114503
	\urlprefix\url{https://doi.org/10.1063/1.4967303}
	
	\bibitem{Pcalpaper-P2000113}
	Bhattacharjee D, Lecoeuche Y, , Karki S, Betzwieser J, Bossilkov V, Kandhasamy
	S, Payne E and Savage R~L 2020 {\em Classical and Quantum Gravity\/} {\bf 38}
	015009 \urlprefix\url{https://doi.org/10.1088/1361-6382/aba9ed}
	
	\bibitem{Pcal-G1501518}
	Karki S, Kissel J, Savage R~L and Goetz E 2015 {aLIGO} {DARM} signal chain
	(calibration subway map) Tech. Rep. LIGO-G1501518 LIGO Laboratory
	\urlprefix\url{https://dcc.ligo.org/LIGO-G1501518-v18/public}
	
	\bibitem{Pcal-T2100067}
	Karki S, , Lecoeuche Y, Bhattacharjee D and Savage R~L 2021 Technical note:
	{O3} {Pcal} {LLO} calibration factors Tech. Rep. LIGO-T2100067 LIGO
	Laboratory \urlprefix\url{https://dcc.ligo.org/LIGO-T2100067/public}
	
	\bibitem{Pcal-T2100219}
	Bhattacharjee D, Karki S, Lecoeuche Y,  and Savage R~L 2021 Technical note:
	{O3} {Pcal} {LHO} calibration factors Tech. Rep. LIGO-T2100219 LIGO
	Laboratory \urlprefix\url{https://dcc.ligo.org/LIGO-T2100219/public}
	
	\bibitem{Brooks2020}
	Brooks A~F {\em et~al.\/} 2021 {\em Appl. Opt.\/} {\bf 60} 4047--4063
	\urlprefix\url{http://ao.osa.org/abstract.cfm?URI=ao-60-13-4047}
	
	\bibitem{carbone2012sensors}
	Carbone L, Aston S, Cutler R, Freise A, Greenhalgh J, Heefner J, Hoyland D,
	Lockerbie N, Lodhia D, Robertson N {\em et~al.\/} 2012 {\em Classical and
		Quantum Gravity\/} {\bf 29} 115005
	
	\bibitem{aston2012update}
	Aston S~M, Barton M~A, Bell A~S, Beveridge N, Bland B, Brummitt A~J, Cagnoli G,
	Cantley C~A, Carbone L, Cumming A~V, Cunningham L, Cutler R~M, Greenhalgh
	R~J~S, Hammond G~D, Haughian K, Hayler T~M, Heptonstall A, Heefner J, Hoyland
	D, Hough J, Jones R, Kissel J~S, Kumar R, Lockerbie N~A, Lodhia D, Martin
	I~W, Murray P~G, O'Dell J, Plissi M~V, Reid S, Romie J, Robertson N~A, Rowan
	S, Shapiro B, Speake C~C, Strain K~A, Tokmakov K~V, Torrie C, van Veggel A~A,
	Vecchio A and Wilmut I 2012 {\em Classical and Quantum Gravity\/} {\bf 29}
	235004 \urlprefix\url{https://doi.org/10.1088/0264-9381/29/23/235004}
	
	\bibitem{LHOalog55399}
	{LHO Electronic Logbook Entry 55399}
	\url{https://alog.ligo-wa.caltech.edu/aLOG/index.php?callRep=55399}
	
	\bibitem{Electronics-G2000527}
	Kissel J 2020 Understanding {H1's} {O3B} electronics compensation systematic
	error Tech. Rep. LIGO-G2000527 LIGO Laboratory
	\urlprefix\url{https://dcc.ligo.org/LIGO-G2000527/public}
	
	\bibitem{Viets2018}
	Viets A~D, Wade M, Urban A~L, Kandhasamy S, Betzwieser J, Brown D~A,
	Burguet-Castell J, Cahillane C, Goetz E, Izumi K, Karki S, Kissel J~S,
	Mendell G, Savage R~L, Siemens X, Tuyenbayev D and Weinstein A~J 2018 {\em
		Classical and Quantum Gravity\/} {\bf 35} 095015
	\urlprefix\url{https://doi.org/10.1088/1361-6382/aab658}
	
	\bibitem{TDCF-T1700106}
	Goetz E, Kandhasamy S, Kissel J, Viets A and Anand S 2019 Update to tracking
	temporal variations in {DARM} loop model parameters: Individual actuation
	stage tracking, cancelled lines, and {SRC} detuning Tech. Rep. LIGO-T1700106
	LIGO Laboratory \urlprefix\url{https://dcc.ligo.org/LIGO-T1700106/public}
	
	\bibitem{BadTDCFs-G2001293}
	Kissel J 2020 {H1's} {OMC} whitening systematic error impact on {TDCF} optical
	gain and cavity pole estimate Tech. Rep. LIGO-G2001293 LIGO Laboratory
	\urlprefix\url{https://dcc.ligo.org/LIGO-G2001293/public}
	
	\bibitem{prokhorov2010space}
	Prokhorov L and Mitrofanov V 2010 {\em Classical and Quantum Gravity\/} {\bf
		27} 225014
	
	\bibitem{GPR}
	Rasmussen C and Williams C 2006 {\em Gaussian Processes for Machine Learning\/}
	Adaptive Computation and Machine Learning (Cambridge, MA, USA: MIT Press)
	
	\bibitem{buhmann2003radial}
	Buhmann M~D 2003 {\em Radial basis functions: theory and implementations\/}
	vol~12 (Cambridge university press)
	
	\bibitem{LHOaLOG55012}
	{LHO Electronic Logbook Entry 55012}
	\url{https://alog.ligo-wa.caltech.edu/aLOG/index.php?callRep=55012}
	
	\bibitem{GPR-length-scale}
	Goetz E, Kissel J and Sun L 2021 Physical meaning of gpr radial basis function
	length scale Tech. Rep. LIGO-G2101319 LIGO Scientific Collaboration
	\urlprefix\url{https://dcc.ligo.org/LIGO-G2101319/public}
	
	\bibitem{Lindblom2009}
	Lindblom L 2009 {\em Phys. Rev. D\/} {\bf 80}(4) 042005
	\urlprefix\url{https://link.aps.org/doi/10.1103/PhysRevD.80.042005}
	
	\bibitem{abbott2016gw150914}
	Abbott B~P, Abbott R, Abbott T, Abernathy M, Acernese F, Ackley K, Adams C,
	Adams T, Addesso P, Adhikari R {\em et~al.\/} 2016 {\em Physical Review D\/}
	{\bf 93} 122003
	
	\bibitem{vitale2020physical}
	Vitale S, Haster C~J, Sun L, Farr B, Goetz E, Kissel J and Cahillane C 2020
	{\em arXiv preprint arXiv:2009.10192\/}
	
	\bibitem{payne2020gravitational}
	Payne E, Talbot C, Lasky P~D, Thrane E and Kissel J~S 2020 {\em Physical Review
		D\/} {\bf 102} 122004
	
	\bibitem{cahillane2017calibration}
	Cahillane C, Betzwieser J, Brown D~A, Goetz E, Hall E~D, Izumi K, Kandhasamy S,
	Karki S, Kissel J~S, Mendell G {\em et~al.\/} 2017 {\em Physical Review D\/}
	{\bf 96} 102001
	
\end{thebibliography}
\end{document}